\documentclass{aa}

\usepackage[varg]{txfonts}
\usepackage{graphicx}
\usepackage{url}
\usepackage{natbib}
\usepackage{color}
\usepackage[normalem]{ulem}
\newcommand{\kms}{km\,${\rm s}^{-1}$}

\include{bibdefinitions}
\begin{document}

\title{The extreme O-type spectroscopic binary HD\,93129A \thanks{Based on observations made with the NASA/ESA Hubble Space Telescope, obtained [from the Data Archive] at the Space Telescope Science Institute, which is operated by the Association of Universities for Research in Astronomy, Inc., under NASA contract NAS 5-26555. These observations are associated with program GO-13346.},\thanks{Based on observations collected at the European Organisation for Astronomical Research in the Southern Hemisphere under ESO programme 095.D-0234(A).}}
\subtitle{A quantitative, multiwavelength analysis}

\author{D. Gruner  \inst{1}
  \and R. Hainich \inst{1}
  \and A. A. C. Sander \inst{2}
  \and T. Shenar \inst{3}
  \and H. Todt \inst{1}
  \and L. M. Oskinova \inst{1}
  \and V. Ramachandran \inst{1}
  \and T. Ayres \inst{4}
  \and W.-R. Hamann \inst{1}}

\institute{Institut f\"ur Physik und Astronomie, Universit\"at Potsdam, Karl-Liebknecht-Str. 24/25, 14476 Potsdam, Germany
\and Armagh Observatory and Planetarium, College Hill, Armagh BT61 9DG, Northern Ireland, UK
\and Instituut voor Sterrenkunde, Universiteit Leuven, Celestijnenlaan 200 D, 3001, Leuven, Belgium
\and Center for Astrophysics and Space Astronomy, 389 UCB, University of Colorado, Boulder, CO 80309, USA
}

\date{Received date /
Accepted date}

\abstract 
  { \object{HD\,93129A} was classified as the earliest O-type star in the Galaxy (O2~If*) and is considered as the prototype of its spectral class. However, interferometry shows  that this object is a binary system, while recent observations even suggest a triple configuration. None of the previous spectral analyses of this object accounted for its multiplicity. With new high-resolution UV and optical spectra, we have the possibility to reanalyze this key object, taking its binary nature into account for the first time.}
  { We aim to derive the fundamental parameters and the evolutionary status of HD\,93129A, identifying the contributions of both components to the composite spectrum}
  {We analyzed UV and optical observations acquired with the Hubble Space Telescope and ESO's Very Large Telescope. A multiwavelength analysis of the system was performed using the latest version of the Potsdam Wolf-Rayet model atmosphere code.}
  {Despite the similar spectral types of the two components, we are able to find signatures from each of the components in the combined spectrum, which allows us to estimate the parameters of both stars. We derive $\log (L/L_\odot) = 6.15$, $T_{\textrm{eff}}=52$\,kK, and log\,$\dot{M}=-4.7\,[M_\odot\text{yr}^{-1}]$ for the primary Aa, and $\log (L/L_\odot)=5.58$, $T_{\textrm{eff}}=45$\,kK, and log\,$\dot{M}=-5.8\,[M_\odot\text{yr}^{-1}]$ for the secondary Ab. } 
  {Even when accounting for the binary nature, the primary of HD\,93129A is found to be one of the hottest and most luminous O stars in our Galaxy.  Based on the theoretical decomposition of the spectra, we assign spectral types O2~If* and O3~III(f*) to components Aa and Ab, respectively. While we achieve a good fit for a wide spectral range, specific spectral features are not fully reproduced. The data are not sufficient to identify contributions from a hypothetical third component in the system.  }


\maketitle

\titlerunning{<short title>}
\authorrunning{<name(s) of
author(s)}

\section{Introduction}
  \label{sec:introduction}

  Hot, massive stars ($\gtrsim10\,M_\odot$) have a strong effect on their surroundings owing to their very high luminosities and powerful stellar winds. Upon death, they may trigger the evolution of new generations of stars through supernova explosions. 

\begin{figure*} \centering
    \includegraphics[scale=0.93,angle=-90]{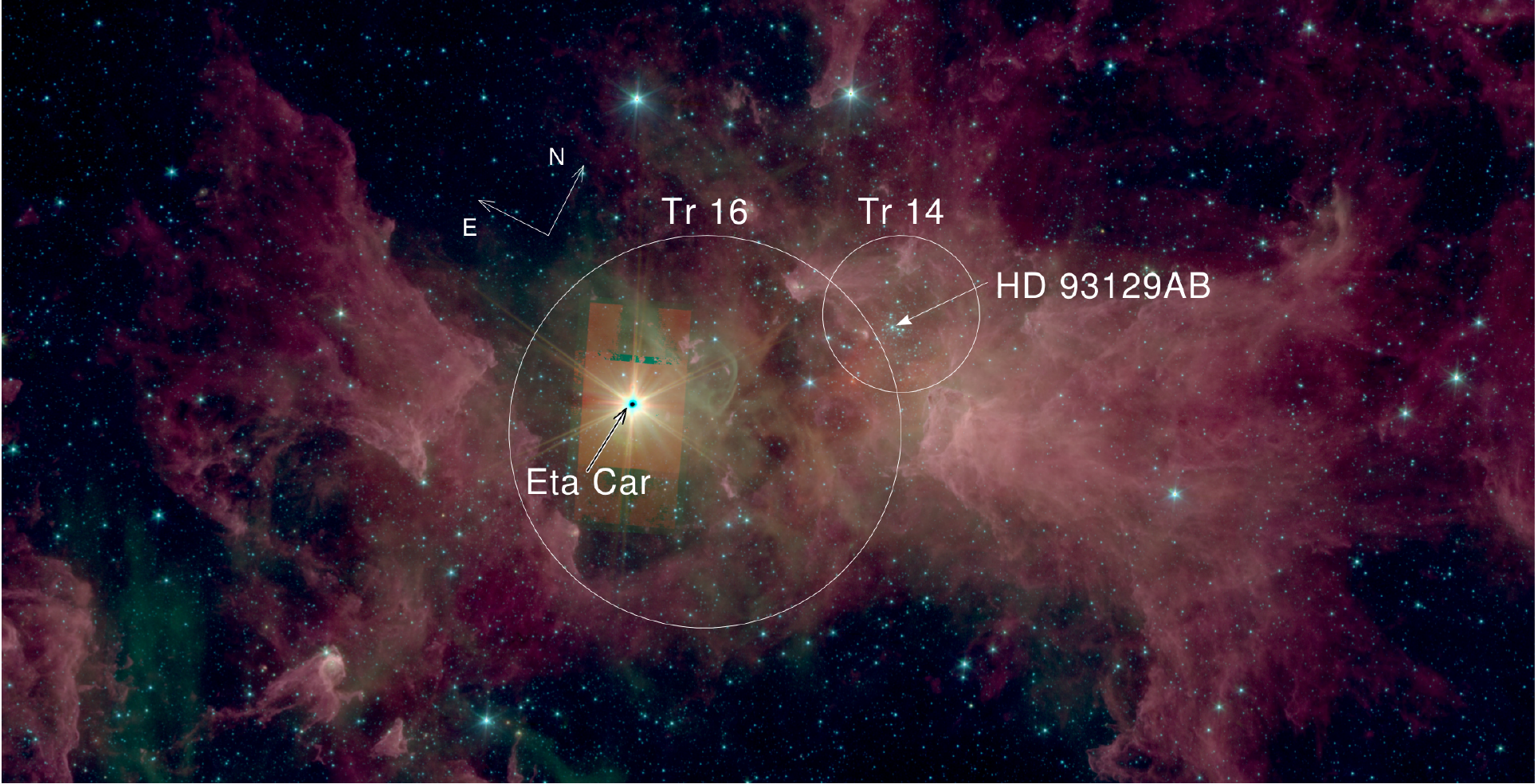} 
    \caption{ Spitzer Image of the Carina Nebula; overlay of IRAC4 at $8\mu$ (red), IRAC2 at $4.5\mu$ (green), and IRAC1 at $3.6\mu$ (blue) on a logarithmic color scale. The dominating stellar clusters Tr 14 and 16 and the two brightest stars Eta Car and HD\,93129AB are indicated. The field size is 66\arcmin$\times 31$\arcmin. \label{fig:carina_full}}
  \end{figure*}

  The Carina Complex (\object{NGC 3372}) (Fig.\,\ref{fig:carina_full}) is dominated by the very bright and enigmatic object \object{$\eta$\,Carinae}, which belongs to  the young cluster \object{Trumpler~(Tr)\,16}. Another cluster in this region is \object{Tr\,14}, harboring our program star \object{HD\,93129A}. These two clusters contain a large number of massive stars and are responsible for the remarkable incandescence of the \object{Carina Nebula}, which is visible to the naked eye despite of its distance of roughly 2.9\,kpc \citep{2012AJ....143...41H}. 

  The former classification for \object{HD\,93129A} as a single star and its very high luminosity have aroused much attention. The object was classified as O3f* \citep{1971ApJ...167L..31W} and later as O3 If* \citep{1973ApJ...179..517W}. \citet{2002AJ....123.2754W} then introduced a new spectral class O2 If* with HD\,93129A as its prototype, thus making it the earliest O-type star known in the Galaxy. Until today, only a few O2 stars have been identified: e.g., \object{30\,Dor\,016} and \object{HDE 269810} in the \object{Large Magellanic Cloud (LMC)}  \citep{2010ApJ...715L..74E}. 

  HD\,93129A was first resolved as a binary by \citet{2004AJ....128..323N}\footnote{Erratum \cite{2010AJ....139.2714N}}, using the Hubble Space Telescope (HST) Fine Guidance Sensor (FGS), who reported a brighter primary and a fainter secondary (\object{HD\,93129Aa} and \object{HD\,93129Ab;}  Aa and Ab thereafter). The binary status was confirmed by \citet{2009AJ....137.3358M} through speckle interferometry, by \citet{2014ApJS..215...15S} using the Precision Integrated-Optics Near-infrared Imaging ExpeRiment (PIONIER) at the Very Large Telescope Interferometer and the Sparse Aperture Interferometric Masks with the Nasmyth Adaptive Optics System (NAOS) Near-Infrared Imager and Spectrograph (CONICA), and by \cite{2015A&A...579A..99B}  with long baseline interferometry. With a brightness contrast of $\Delta m_V=0.9$\,mag that has been confirmed multiple times in previous observations, the secondary star provides about one-third of the overall flux in the optical. Four faint nearby stars were found in the immediate vicinity of HD\,93129A: HD\,93129C \citep{2009AJ....137.3358M,2010A&A...515A..26S}, E, F, and G \citep{2014ApJS..215...15S}. By reanalyzing older HST FGS observations,  \cite{2015A&A...579A..99B} made a first estimation of the binary orbit. It was found to be very eccentric and/or seen under high inclination. The orbital period is estimated to be $>100$\,yr. \cite{2016arXiv160407294D} modeled the colliding wind region of the binary and suggested a low orbital inclination of $i<30^\circ$. Interferometry shows that both stars are approaching each other. The most recently derived separation between Aa and Ab is 26.5\,mas \citep{2014ApJS..215...15S}, which corresponds to a separation of 78.2\,AU\ at a distance of 2.9\,kpc \citep{2012AJ....143...41H}. The periastron passage was predicted to occur in late 2017 or early 2018 \citep{2015A&A...579A..99B,2017MNRAS.464.3561M}. 

  Based on early non-LTE model atmospheres, \citet{1983A&A...125...34S} analyzed a spectrum of HD\,93129A by fitting several optical lines. The most extensive analysis so far has been published by \citet{1997A&A...321..531T}. Using \emph{International UV Explorer} (IUE) and \emph{Orbiting and Retrievable Far and Extreme Ultraviolet Spectrometer} (ORFEUS) spectra, they analyzed a wide wavelength range in the (far-) UV together with selected optical lines. The authors of both analyses were not yet aware of the binary nature of this object. Even the most recent analysis by \citet{2004A&A...415..349R} did not account for its binarity.

  Using sophisticated extraction techniques, \citet{2017MNRAS.464.3561M} disentangled the spectra of the two components in an optical HST observation.  An investigation of the spectra made them realize that the primary itself might have two components. We critically discuss this hypothesis in Sect.\,\ref{sec:triple}. Given the uncertain nature of the third component, a three-component analysis cannot be performed unambiguously with our data (see Sect. 6.1). We therefore treat the system as a binary in our study.

  We perform a first consistent spectroscopic non-local thermal equilibrium (LTE) analysis of the two components of the binary system. We use the Potsdam Wolf-Rayet model atmosphere code (PoWR), which is suitable for any type of hot stars, such as OB-stars, luminous blue variables, Wolf-Rayet stars, and central stars of planetary nebula  \citep[e.g.,][]{2003A&A...410..993H}.

  Two issues make the analysis of this system especially difficult: 
    (1) Following the classification of \citet{2004AJ....128..323N}, the two binary components are early O-type stars. Therefore, their spectra look similar. This is most likely the reason why previous analyses by \citet{1983A&A...125...34S} and \citet{1997A&A...321..531T} failed to recognize the binary character of our object. 
    (2) The cluster Tr\,14, like the whole Carina Complex, is densely populated with early-type stars. At a distance of roughly 3\,kpc, the angular separation of the cluster members is small. We show that observations from the \emph{Far Ultraviolet Spectroscopic} (FUSE), IUE, and ORFEUS \citep[the latter two were used by][]{1997A&A...321..531T} suffer from source contamination. 

  High-resolution data with a high signal-to-noise ratio (S/N) form the basis and motivation of our new analysis. High-resolution spectra have been obtained for the UV with the HST, and for the optical and near-infrared (NIR) with the ESO Very Large Telescope (VLT).

  In Sect.\,\ref{sec:observations} we describe and discuss the available observational data. Our modeling methods are described in Sect.\,\ref{sec:powr}. The analysis is presented in Sect.\,\ref{sec:analysis}, and the results are summarized in Sect.\,\ref{sec:results}. The evolutionary status of the system is discussed in the final Sect.\,\ref{sec:discussion}, including the implications of recently published new observations that suggest a third component in the system \citep{2017MNRAS.464.3561M}.

\section{Observational data} \label{sec:observations}

\begin{figure}
    \centering
    \includegraphics[width=1\linewidth]{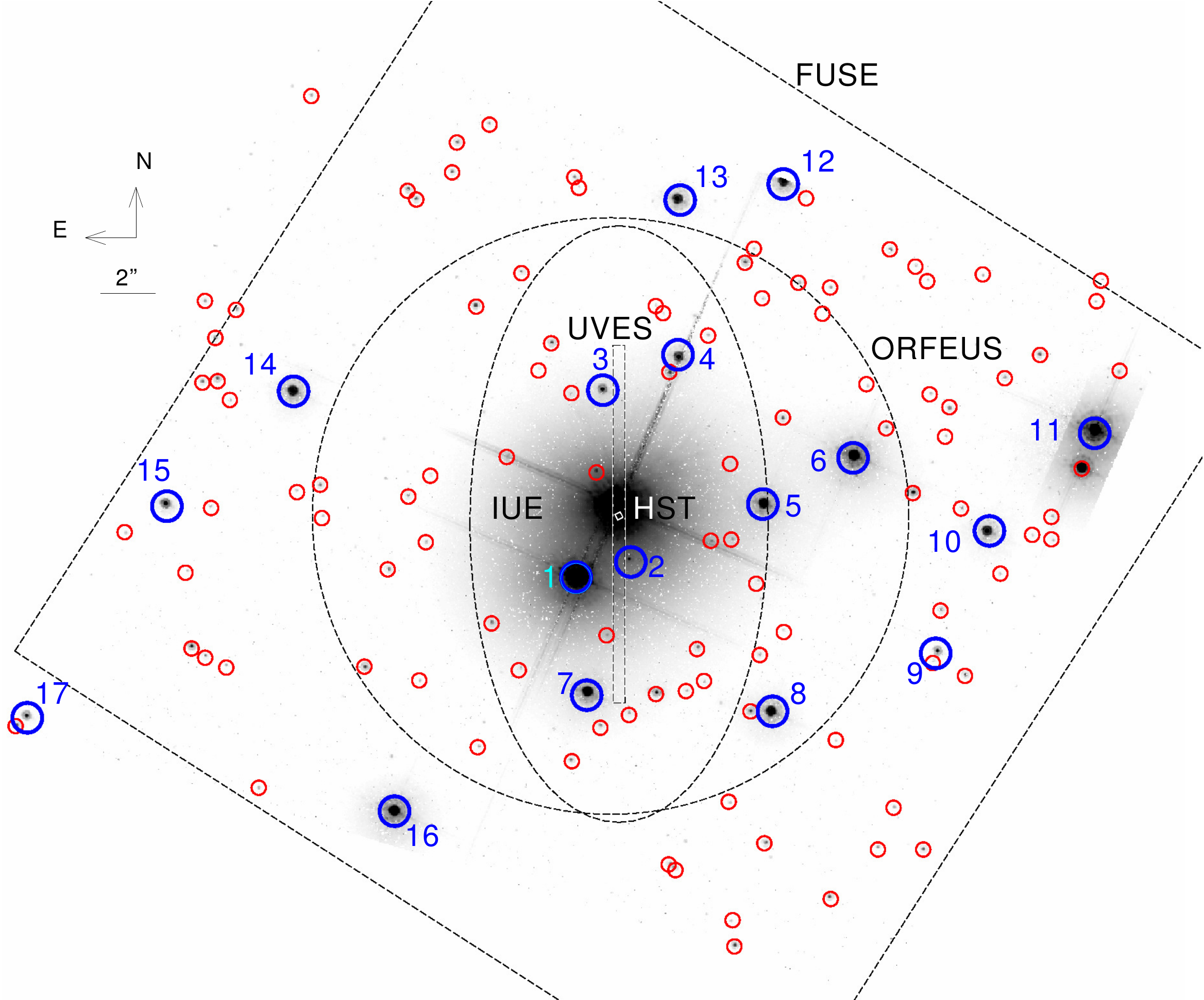} 
    \caption{ Inner region of \object{Tr 14} centered on HD\,93129A. The dashed boxes and ovals indicate the apertures of several instruments that were used for spectroscopic observations. The figure accounts for pointing and position angle of the individual observations, except for IUE, where multiple observations were carried out. The dashed ellipse shows that the IUE large aperture adopted a perfect pointing and a position angle of zero. The blue circles identify previously cataloged objects (Table \ref{tab:cluster_stars}). Red circles mark further objects that are not listed in catalogs so far. Background image: HST ACS/HRC F850LP on a logarithmic and inverted grayscale. \label{fig:bigapertures}} 
  \end{figure}

  The previously available pre-HST UV-data were recorded with IUE, FUSE, and ORFEUS. Figure\,\ref{fig:bigapertures} shows the inner region of Tr\,14 centered on HD\,93129A. It is evident that \object{HD\,93129B} \citep[an O3.5V((f)) star,][]{2014ApJS..211...10S} contaminated all previous observations\footnote{This even holds for the IUE small aperture, which is not shown in the figure}. Other identified stars in the region that likely contaminate the observations are listed in Table \ref{tab:cluster_stars}. Thus all available data contain at least three early type O-stars. 

  \begin{table}
    \centering
    \caption{Stars in the central region of \object{Tr 14}. The numbers in the first column correspond to the labels in Fig.\,\ref{fig:bigapertures}.  \label{tab:cluster_stars}}
    \begin{tabular}{ r l ccc c }
      \hline\hline\\[-2.0ex]
      \# & Star  & U & B & V & Ref.                             \\
      \hline\\[-2.0ex]
        & \object{HD\,93129A}                        & 6.70     & 7.51  & 7.26    & 2 \\
        & \object{HD\,93129Aa}                       & 7.09     & 7.90  & 7.65    &  8 \\   
        & \object{HD\,93129Ab}                       & 7.99     & 8.80  & 8.55    & 8 \\
      \hline\\[-2.0ex]
      1 & \object{HD\,93129B}                        &-         & 10.4  &8.8    &   \\   
      2 & \object{HD\,93129E}                        & -        & -     & -       & 7 \\
      3 & \object{HD\,93129F}                        & -        & -     & -       & 7 \\
      4 & \object{HD\,93129G}                        & -        & -     & -       & 7\\
      5 & \object{HD\,93129C}                        & -        & -     & -       & 3\\
      6 & \object{VBF 17}                            &10.4      &11.1   &10.9   & 2 \\
      7 & \object{VBF 36}                            &- &11.9   &12.1   & 2 \\
      8 & \object{VBF 63}                            &13.1      &13.1   &12.7   & 2\\
      9 & \object{$[$HSB2012$]$ 1508}                & -        & -     &16.0   & 5  \\
      10  & \object{$[$HSB2012$]$ 1498}              &12.5      &13.1   &12.8   & 6 \\     
      11  & \object{Tr 14 9}                         &9.3       &10.1   &9.9    & 1\\
      12  & \object{MJ 173}                          &12.6      &13.0   &12.8   & 1\\
      13  & \object{VBF 65}                          &12.9      &13.2   &12.9   & 2\\
      14  & \object{MJ 186}                          &12.3      &12.8   &12.5   & 1\\
      15  & \object{VBF 100}                         &14.2      &14.4   &13.8   & 2\\
      16  & \object{Tr 14 19}                        &11.8      &12.0   &11.7   & 1\\
      17  & \object{$[$HSB2012$]$ 1650}              & -        &18.2   &16.6   & 4\\
      \hline
    \end{tabular}
    \tablebib{ Objects firstly discovered by: 
      (1) \cite{1993AJ....105..980M}; (2) \cite{1996A&AS..116...75V}; (3) \cite{1998AJ....115..821M}; (4) \cite{2001ApJ...549..578D}; (5) \cite{2003yCat.2246....0C}; (6) \cite{2012AJ....143...41H}; (7) \cite{2014ApJS..215...15S}, (8) \citep{2004AJ....128..323N};\\ Photometry of HD\,93129Aa and Ab are derived here.}
  \end{table}

\subsection{UV spectrum}

  The Advanced SpecTRAl Library program \citep[ASTRAL,][]{Ayres.2015}, an HST Large Treasury Project, was designed for the creation of an atlas of contiguous high-resolution UV-spectra using the Space Telescope Imaging Spectrograph (STIS). Twenty-one early-type stars were selected as representatives for their spectral class\footnote{\label{fn:astral}see \url{http://casa.colorado.edu/~ayres/ASTRAL/}}. The characteristics and operational capabilities of the HST have been described in detail, e.g., \citet{1998PASP..110.1183W} and \citet{1998ApJ...492L..83K}.

  To obtain a contiguous UV spectrum for HD\,93129A, the complete UV wavelength range was covered by three overlapping spectograph settings (see Table \ref{tab:obs_overview}). Each range was exposed twice. These six observation were carried out on 23 October 2014 (visit~1) and 6 November 2014 (visit~2).

  \begin{table*}
    \centering
    \caption{Observation datasets from the ASTRAL project for HD\,93129A  \label{tab:obs_overview}} 
    \begin{tabular}{ c c c c c c c c c c  }
        \hline\hline
        \rule[0mm]{0mm}{4mm}
        Dataset                                         &  Aperture   & Grating & Position Angle    & Wavelength Range  & $t_{\text{Exp}}$ &  Epoch  \\
                                                        &    [\arcsec]     &     &   [${}^\circ$]       &  [\AA]                 &        [s]       &     [MJD]       \\
        \hline
        ocb6a0020 \rule[0mm]{0mm}{4mm}  & 0.20$\times$0.09 & E140M-1425 & 191 & 1140-1709             & 2636 &  56953.8    \\
        ocb6a1020                       & 0.20$\times$0.09 & E140M-1425 & 207 & 1140-1709             & 2636 &  56967.8     \\
        ocb6a0010                       & 0.20$\times$0.20 & E230M-1978 & 191 &  1610-2365        & 1729 &  56953.8    \\
        ocb6a1010                       & 0.20$\times$0.20 & E230M-1978 & 207 & 1610-2365             & 1729 &  56967.8 \\
        ocb6a0030                       & 0.10$\times$0.03 & E230M-2707 & 191 &  2280-3071        & 1900 &  56953.8      \\
        ocb6a1030                       & 0.10$\times$0.03 & E230M-2707 & 207 &   2280-3071       & 1900 &  56967.8   \\
        \hline
    \end{tabular}
  \end{table*}

\begin{figure}\centering
    \includegraphics[scale=1, angle=-90]{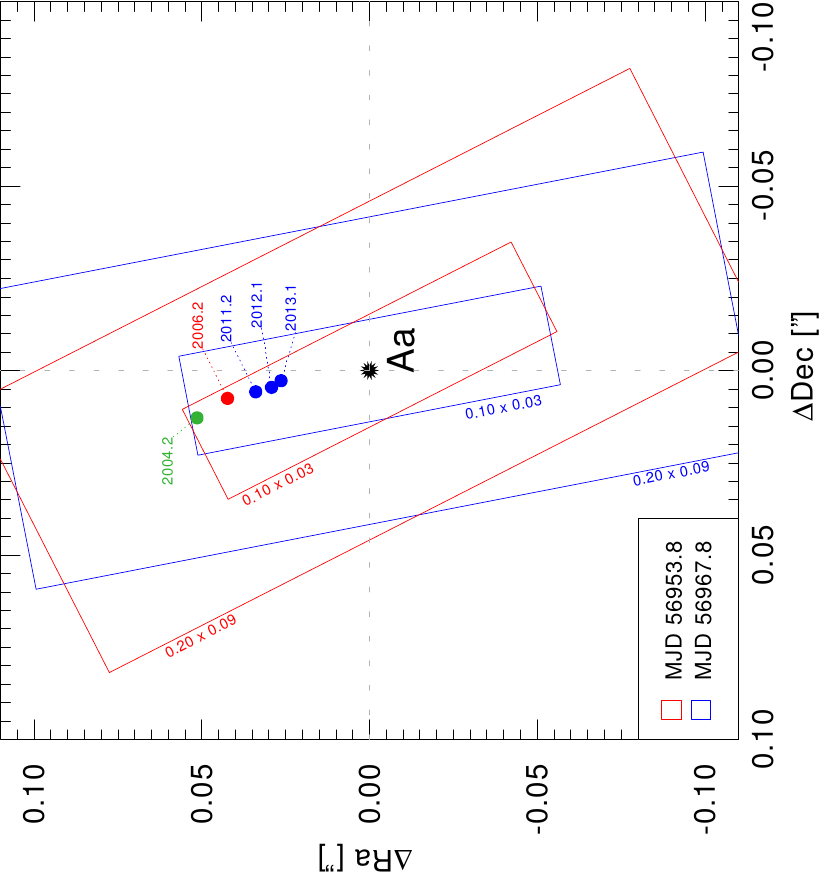} 
    \caption{ STIS slits of the observations compared to the relative positions of the binary components. Only the two smaller slit configurations from Table \ref{tab:obs_overview} are shown for both pointings. The position of Ab at different epochs is indicated by filled circles (blue: \citet{2014ApJS..215...15S},  red: \cite{2009AJ....137.3358M}, green: \citet{2010AJ....139.2714N}). The position of Ab may be extrapolated to the epoch of our observation (2014.8), and is obviously covered by even the smallest STIS slit. The centering procedure with STIS would peak on the center of (visible) light of the binary, rather than placing HD\,93129Aa exactly at (0,0).  This strengthens the argument that Aa and Ab were both entirely covered by all the STIS apertures. \label{fig:smallapertures_up}}      
  \end{figure}

  In Fig.\,\ref{fig:bigapertures} the largest STIS slit configuration used is marked in the center, covering only HD\,93129A. Figure \ref{fig:smallapertures_up} shows the used STIS apertures, assuming a perfect centering on the position of Aa. The measured locations of Ab at different epochs are indicated by filled circles. A clear motion that has been described by \citet{2015A&A...579A..99B} is visible, but the extrapolated position at the epoch of our observation (2014.8) is  covered by even the smallest STIS slit.

  The STIS echellograms of HD\,93129A were initially passed through the standard CALSTIS pipeline.  The resulting files then were post-processed to (1) correct the wavelength scale for small errors in the default dispersion relations; (2) shift the echelle sensitivity (``blaze'') to achieve the best global match between fluxes in the overlapping zones of adjacent orders, and (3) merge the echelle orders to provide a coherent 1D spectral tracing for the specific grating setting.  Next followed a series of steps to merge the independent exposures of a given setting, and finally splice the three partially overlapping wavelength regions into a full-coverage UV spectrum.  A bootstrapping approach, applied to the overlaps, optimized the precision of the global relative wavelength scale and stellar energy distribution.  A detailed description of the ASTRAL protocols can be found on the ASTRAL web site (footnote\,\ref{fn:astral}).

\subsection{Optical spectrum}

  \object{HD\,93129A} was observed with the UV to Visual Echelle Spectograph (UVES) mounted at the Kuyen 8\,m telescope of the ESO VLT at the Paranal observatory on the night of 17-18\ April 2015. To obtain the full wavelength coverage and the highest spectral resolution, the two standard settings DIC1 346+580 and DIC2 437+860 were used, with a 0\farcs4 slit for the blue and a 0\farcs3 slit for the red arm, and slit lengths of $12\arcsec$ and $8\arcsec$ , respectively. The resolution of the spectra ranges from 80\,000 in the blue arm to 115\,000 in the red arm. The spectra were reduced with the UVES pipeline data reduction software. Using an exposure time of 12\,min for the blue arm and 10\,min for the red arm, we  reached an S/N of about 200 and 250 was reached, respectively. The spectra are not flux calibrated and were rectified by fitting a low-order polynomial to the apparent continuum.

  In Fig.\,\ref{fig:bigapertures} we indicate the larger of the two UVES slit configuration (12$\arcsec\times$0\farcs4). The used positioning angle was zero for the two exposures and therefore only covered HD\,93129A, but E and F are close to the edge of the slit. According to the ESO Observatories Ambient Conditions Database, the mean seeing for the observation night was $\approx$1\farcs5, implying that both stars  likely contaminated the observations. To evaluate the effect that this had, we consulted the interferometric observations obtained by \cite{2009AJ....137.3358M}. These authors were unable to resolve stars E and F down to a magnitude difference of $\Delta m_V < 3.5$\,mag compared to HD\,93129A. We can therefore safely assume that the contamination caused by these stars is negligible.

\subsection{Photometry}
  \label{sec:photometry}

  For UBVRI photometry, we adopted the measurements from \cite{1996A&AS..116...75V}. Their observations were obtained with a spatial resolution of $0\farcs45/$pixel and a mean seeing of $1\farcs3$ and should therefore minimize the contamination from nearby sources. Comparison of their B- and V-band magnitudes to the Tycho-2 Catalog \citep{2000A&A...355L..27H} taken with HIPPARCOS reveals good agreement.  Furthermore, the measurements are consistent with GAIA DR2 photometry \citep{2018arXiv180409368E}.

  The UV photometry was extracted from two HST ACS/HRC images taken with the F220W and F435W\footnote{Observation ID: \url{hst_10205_01_acs_hrc_f220w} and \url{hst_10205_01_acs_hrc_f435w}, respectively} filter. The data were obtained with MAST from the Hubble Legacy Archive.

  The JHK photometry is taken from the 2MASS point source catalog \citep{2006AJ....131.1163S}. We note that the spatial resolution of 2MASS is too low for this crowded region. Contamination of HD\,93129B, further nearby sources, and the Carina Nebula itself cannot be neglected.  The JHK photometry is therefore expected to overestimate the flux (see Fig.\,\ref{fig:sed_compare}). The photometry we used for our analysis is summarized in Table\,\ref{tab:photometry}.
  
  \begin{table}
    \centering
    \caption{Photometric measurements of HD\,93129A \label{tab:photometry}}
    \begin{tabular}{rcc}
      \hline\hline\\[-2.0ex]
      Band  & Magnitude  & Ref.                             \\
      \hline\\[-2.0ex]
      HST ACS F220W  & 8.92  &    1 \\   
      HST ACS F435W  & 8.18  & 1 \\
      Johnson U      & 6.70  & 2 \\   
      Johnson B      & 7.51  & 2 \\
      Johnson V      & 7.26  & 2 \\   
      Johnson R      & 7.08  & 2 \\
      Johnson I      & 6.82  & 2 \\   
      GAIA G         & 7.16  & 3 \\
      2MASS J        & 6.23  & 4 \\   
      2MASS H        & 6.14  & 4 \\
      2MASS K        & 6.01  & 4 \\
      \hline
    \end{tabular}
    \tablebib{ Photometry taken from 
      (1) this work; (2) \cite{1996A&AS..116...75V}; (3) \cite{2018arXiv180409368E}; (4) \cite{2006AJ....131.1163S}.}
  \end{table}

\section{PoWR model atmospheres} 
  \label{sec:powr}

  To analyze the observed spectra, we fit them with synthetic spectra calculated with the Potsdam Wolf-Rayet (PoWR) model atmospheres code. The PoWR code solves the non-LTE comoving frame radiative transfer equations for a spherically expanding atmosphere simultaneously with the statistical equilibrium equations. At the same time, it accounts  for energy conservation. More details of the code can be found in \citet{1998A&A...335.1003H}, \citet{2002A&A...387..244G}, and \citet{2015A&A...577A..13S}. 

  The inner boundary of the model atmosphere is set at a Rosseland-mean continuum optical depth of $\tau_{\textrm{Ross}}=20$ and defines the stellar radius $R_*$. The stellar temperature $T_*$ refers to $R_*$ and the    luminosity $L$ via the Stefan-Boltzmann-law. Since the models for this paper are for winds that are optically thin in the continuum, $T_*$ is almost identical to $T_{2/3}$, which refers to the radius where the Rosseland optical depth has dropped to two-thirds.  

  For the expanding part of the atmosphere, the velocity field is prescribed in the form of a $\beta$-law, where the exponent $\beta$ and the terminal velocity $v_\infty$ are free parameters. The mass-loss rate $\dot M$  that we determine is another free parameter. Together with the velocity field, it defines the density stratification through the equation of continuity. In the subsonic part of the atmosphere, the density stratification is modeled such that it approaches hydrostatic equilibrium. When solving the hydrostatic equation, we take special care to account for the radiation pressure consistently \citep{2015A&A...577A..13S}. The surface gravity $\log g$ is a fundamental parameter for the quasi-hydrostatic part of the model atmosphere.   

  Microturbulence also has some impact on the density structure (through turbulence pressure), but it affects the line formation in particular. In the formal integral, we specify the micoturbulent velocity by its minimum value  $v_{\rm mic}$ in the photosphere and a contribution that grows proportional to the wind velocity \citep[see also][]{2015ApJ...809..135S}.  

  Wind inhomogeneities are accounted for in the microclumping approximation, that is,\ under the assumption that all clumps are optically thin at all frequencies. The density in the clumps is enhanced by a factor $D$ compared to a homogeneous wind with the same mass-loss rate, while the interclump space is assumed to be void. There are indications that clumps exist already at or very close to the photosphere \citep{2009A&A...499..279C,2015ApJ...810..102T}. However, we assume that the clumping factor grows from 1 (unclumped) at the photosphere and reaches its maximum value $D$ at $v = 1/4\, v_\infty$.

  Photospheric absorption lines show pressure-broadened profiles. Fitting their wings is the main diagnostic for the star's gravity $\log g$. Pressure broadening is taken into account in the formal integral of the radiative transfer equation,  which yields the emergent spectrum in the observer's frame. 

  The rotational broadening of photospheric lines can be simulated by convolving the emergent flux spectrum with a semi-ellipse.  For lines formed in the stellar wind, however, this is a poor approximation. Therefore, we performed a full 3D integration \citep{2014A&A...562A.118S}, assuming co-rotation of the photosphere ($\tau_\text{Ross} > 2/3$) and conservation of angular momentum in the wind. 

  The models calculated for this paper account for complex model atoms of H, He, C, N, O, Si, P, and S. Iron-group elements are treated with the superlevel approximation \citep{2002A&A...387..244G}. A list of the included ions as well as the number of accounted levels and line transitions is given in Table\,\ref{tab:powr_lvl}.

  \begin{table}[h]\centering\
    \caption{Numbers of atomic levels and transitions that are accounted for in the model calculation. The numbers for iron refer to superlevels and superlines; Fe stands for a generic element that includes the whole iron group. \label{tab:powr_lvl}}
    \begin{tabular}{lrr|lrr}
      \hline
      \hline
      Ion    \rule[0mm]{0mm}{4mm} &  Levels  & Transitions & Ion &  Levels  & Transitions  \\
      \hline
      \ion{H}{   i}\rule[0mm]{0mm}{4mm}    &  22  &  231   &           \ion{Si}{  iv}    &  27  &  351 \\
      \ion{H}{  ii}    &   1  &    0   & \ion{Si}{   v}    &  11  &   55 \\
      \ion{He}{   i}    &  35  &  595   &    \ion{Si}{  vi}    &   1  &    0 \\
      \ion{He}{  ii}    &  26  &  325   &           \ion{P}{  iv}    &  12  &   66 \\
      \ion{He}{ iii}    &   1  &    0   &           \ion{P}{   v}    &  11  &   55 \\
      \ion{C}{  ii}    &  32  &  496   &       \ion{P}{  vi}    &   1  &    0 \\
      \ion{C}{ iii}    &  40  &  780   &       \ion{S}{ iii}    &   1  &    0 \\
      \ion{C}{  iv}    &  25  &  300   &       \ion{S}{  iv}    &  11  &   55 \\
      \ion{C}{   v}    &  29  &  406   &       \ion{S}{   v}    &  10  &   45 \\
      \ion{C}{  vi}    &   1  &    0   &       \ion{S}{  vi}    &  22  &  231 \\
      \ion{N}{  ii}    &  38  &  703   &         \ion{Fe}{ iii}    &   1  &    0 \\
      \ion{N}{ iii}    &  56  &  540   &         \ion{Fe}{  iv}    &  18  &   77 \\   
      \ion{N}{  iv}    &  38  &  703   &         \ion{Fe}{   v}    &  22  &  107 \\   
      \ion{N}{   v}    &  20  &  190   &         \ion{Fe}{  vi}    &  29  &  194 \\   
      \ion{N}{  vi}    &  14  &   91   &         \ion{Fe}{ viii}    &  19  &   87 \\   
      \ion{N}{ vii}    &   2  &    1   &         \ion{Fe}{viii}    &  14  &   49 \\   
      \ion{O}{  ii}    &  37  &  666   &            \ion{Fe}{  iv}    &  15  &   56 \\
      \ion{O}{ iii}    &  33  &  528   &            \ion{Fe}{   x}    &  28  &  170 \\
      \ion{O}{  iv}    &  29  &  406   &            \ion{Fe}{  xi}    &  26  &  161 \\
      \ion{O}{   v}    &  36  &  630   &            \ion{Fe}{ xii}    &  13  &   37 \\
      \ion{O}{  vi}    &  16  &  120   &            \ion{Fe}{xiii}    &  15  &   50 \\
      \ion{O}{ vii}    &   1  &    0   &            \ion{Fe}{ xiv}    &  14  &   49 \\
      \ion{Si}{ iii}    &  24  &  276   &           \ion{Fe}{  xv}    &   1  &    0 \\ 
      \hline
    \end{tabular}
  \end{table}

\section{Analysis} \label{sec:analysis}

  \begin{figure*}
    \centering
    \includegraphics[angle=-90, width = 1\linewidth]{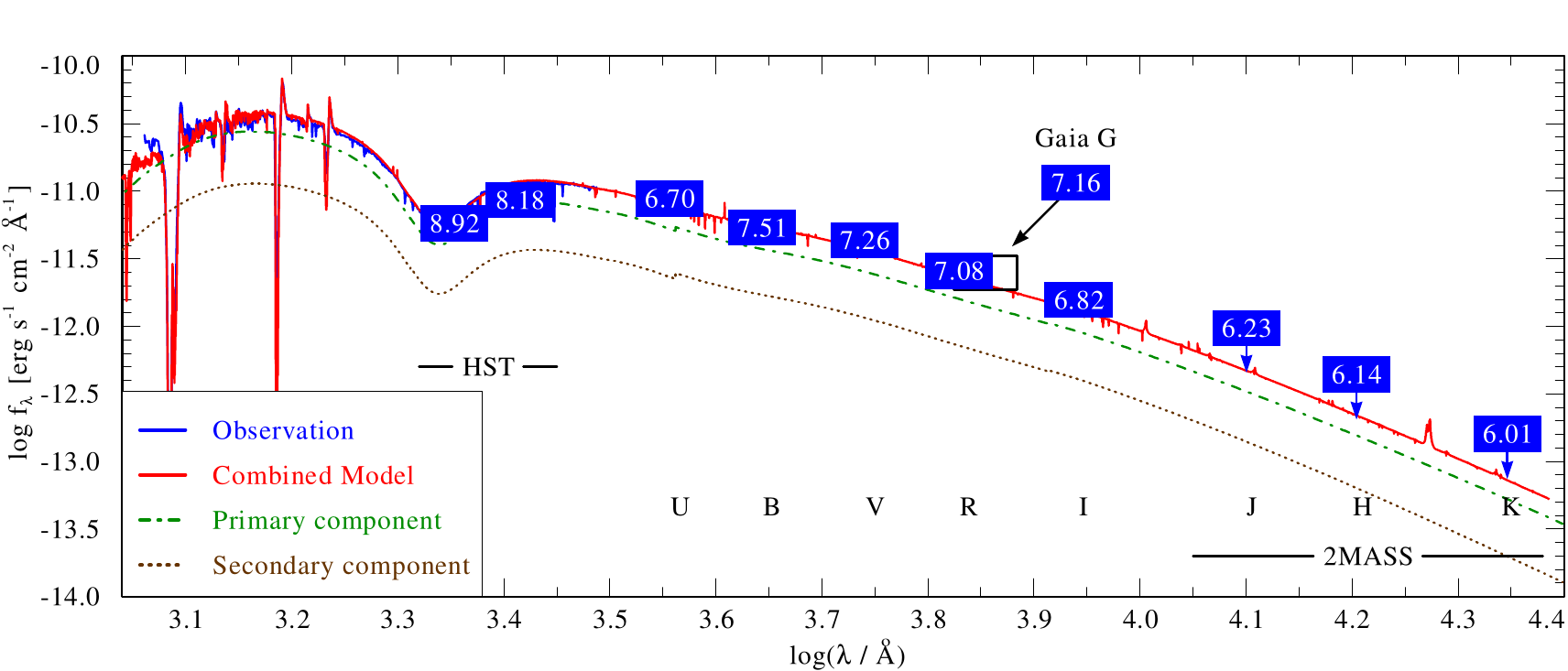} 
    \caption{ Comparison of the observed flux-calibrated UV spectrum (blue solid line) and the photometry (blue boxes with magnitudes imprinted) to the combined, diluted, and reddened synthetic SED (red solid line). The SEDs of the individual components are indicated by the green dash-dotted line (primary) and the brown dotted line (secondary), respectively. The excess in the IR is due to contamination of the 2MASS photometry by nearby stars, and thus only provides upper limits. The Gaia G and the Johnson R photometry overlap. For visibility, the Gaia photometry box is shifted. The black box and arrow indicate its actual position. \label{fig:sed_compare}}
  \end{figure*}

 The analysis of binaries typically takes advantage of Doppler shifts that are caused by the orbital motion, which can be used to distinguish between or disentangle the components in a time series of observations \citep[e.g.,][]{2017A&A...598A..84A, 2017A&A...598A..85S}. However, because of the large separation between Aa and Ab and the long orbital period, their radial velocities hardly changed over the time covered by observations. The primary and secondary both have early spectral types \citep[O2 and O3.5, as assumed by][]{2004AJ....128..323N}, which means that they show similar spectra. If we wish to perform a consistent analysis that fully accounts for the binary character, we need to identify spectral features that can be unambiguously attributed to the individual components. For an initial guess on the stellar parameters, we adopted the results from \citet{1997A&A...321..531T} for the primary and the calibration from \cite{2005A&A...436.1049M} for the secondary based on its spectral type. 
 
\subsection{Brightness ratio and absolute magnitudes}

  A key ingredient for the analysis and the decomposition of the individual components spectra is the brightness ratio. With the V-band photometry $V=7.3$\,mag and the V-band brightness ratio, given as $\Delta m_V=0.9$\,mag \citep{2004AJ....128..323N}, we can calculate the component brightnesses. \citet{2012AJ....143...41H} used main-sequence fitting to derive distance and reddening toward Tr\,14. They derived a distance of $d=2.9$\,kpc, corresponding to a distance modulus $\mu=12.3$\,mag. 

  This distance agrees well with the parallax measured by Gaia DR2 \citep{2016A&A...595A...1G,2018arXiv180409365G,2018arXiv180409376L}\footnote{GAIA source id of HD\,93129A: 5350363910256783488} , which corresponds to a distance of $d=2.8^{+0.3}_{-0.2}$\,kpc  \citep{2018arXiv180410121B}. 

  For the interstellar reddening, \citet{2012AJ....143...41H} suggested two components. One part accounts for the interstellar medium (ISM) with $E_{B-V}^{\text{ISM}}=0.36$\,mag. For this foreground extinction, we adopted  a total-to-selective extinction ratio of $R_{V}^{\text{ISM}}=3.0$ as typical for  the solar neighborhood \citep{1989AJ.....98..611G}. The second part is an intracluster component with a total-to-selective extinction ratio of $R_{V}^{\text{cluster}}=4.4$. The cluster reddening $E_{B-V}^{\text{cluster}}$ is treated as a free parameter.

  To determine $E_{B-V}^{\text{cluster}}$, we calculated stellar atmosphere models and combined the spectral energy distributions (SED) of the components. PoWR provides us with absolute fluxes for the spectra.  These fluxes were then diluted according to $\mu$ and reddened using  the composed reddening law as described above. At this point, we have three free parameters, the individual component luminosities and $E_{B-V}^{\text{cluster}}$. The binary total brightness and the brightness ratio place constraints on the luminosities. The flux-calibrated UV spectrum determines the reddening parameter. We fit the combined SED to the photometry and the flux-calibrated HST UV spectrum, as demonstrated in Fig.\,\ref{fig:sed_compare}. We tested the reddening laws by \citet{1979MNRAS.187P..73S}, \citet{1989ApJ...345..245C}, and \cite{1999PASP..111...63F}. The best fit was achieved for an inner cluster extinction of $E_{B-V}^{\text{cluster}}=0.1$\,mag using the law from \citet{1989ApJ...345..245C} for both reddening domains. This results in an overall extinction of $E_{B-V}=0.46$\,mag and is inconsistent with the value of $0.54$\,mag suggested by \citet{1973ApJ...179..517W}.

  When our results are applied to the photometry, we find an absolute magnitude of $M_V  = -6.5$\,mag for the binary system. With a magnitude difference $\Delta m_V=0.9$\,mag, we obtain $M_{V}=-6.1$\,mag for the brighter component, Aa, and $M_{V}=-5.2$\,mag for the fainter component, Ab. The model fits are constrained thus that the $V$ band magnitudes of the synthetic spectra match the observed brightness ratio $\Delta m_V=0.9$\,mag. 

  A comparison of the derived absolute magnitudes with the calibrations by \cite{2005A&A...436.1049M} and \cite{2006A&A...457..637M} suggests the spectral type O2~IV for the primary and O5~V for the secondary.  If we were to adopt the spectral types as assumed by \citet{2004AJ....128..323N} (O2\,I+O3.5\,V), the absolute visual brightnesses of the primary and secondary would be $-6.45$\,mag and $-6.1$\,mag, respectively. These values are inconsistent with the observations as the total system brightness would add up to $-7.1$\,mag, which is too bright by a factor of $\approx 1.5$ and shows a too small $\Delta m_V$ of $0.35$. The spectral types we eventually derived from our spectral analysis and the theoretical disentanglement are O2\,I and O3\,III. These spectral types correspond to absolute brightnesses of $-6.44$ and $-6.12$ \citep{2006A&A...457..637M} for the primary and the secondary, respectively. These values are similar to those for the spectral types given by \citet{2004AJ....128..323N} and suffer from the same shortcomings.

  In the rest of this section, we consider the normalized line spectrum in detail. The composite synthetic spectra are normalized to the sum of the model continua from the components. 

\subsection{Surface gravity}

\begin{figure}
    \centering
    \includegraphics[scale=1,angle=-90]{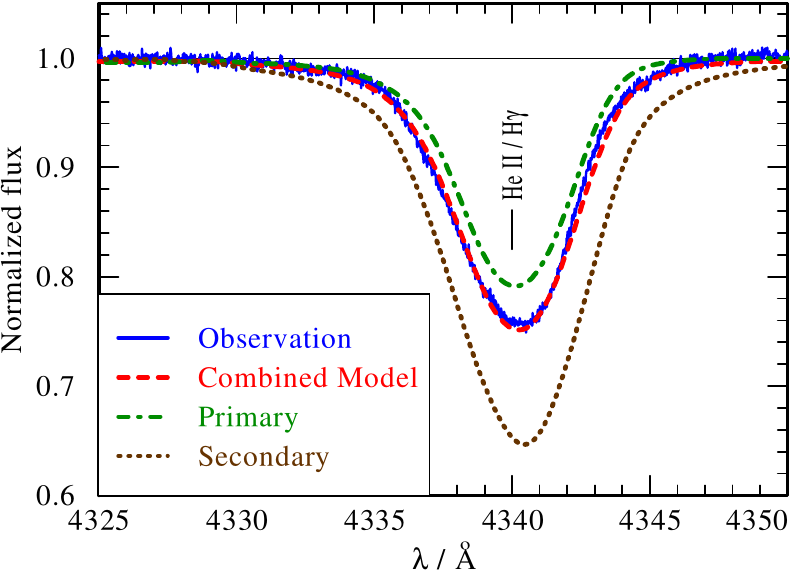} 
    \caption{Observed H$\gamma$ $\lambda 4340$ and \ion{He}{ii} $\lambda 4339$ blend (blue solid line) compared to a combined and normalized spectrum (red dashed line) with log\,$g=4.1$\,[cm\,s$^{-2}$] and log\,$g=3.9$\,[cm\,s$^{-2}$] for Aa (green dash-dotted line) and Ab (brown dotted line), respectively.  \label{pic:logg}}
  \end{figure}

  Usually, $\log g_*$ is derived from the wings of absorption lines belonging to the hydrogen Balmer series. As both components contribute to these lines, an independent determination is impossible. Within limits, raising the gravity for one star can be compensated for by reducing it for the other one to obtain an almost identical fit. Based on the spectral type O3\,V \citep{2004AJ....128..323N}, we assumed a typical gravity of log\,$g=3.9$\,[cm\,s$^{-2}$] for Ab \citep{2005A&A...436.1049M} and adjusted the gravity of Aa such that it reproduced the observation  (Fig.\,\ref{pic:logg}), resulting in log\,$g=4.1$\,[cm\,s$^{-2}$] for Aa. However, given the poor sensitivity of the fit to $\log g_*$, we can only constrain the surface gravity to  $3.7<\textrm{log\,}g<4.2$ for Aa and to $3.8<\textrm{log\,}g<4.2$ for Ab.  

\subsection{Temperature and luminosity}

  The main diagnostics generally used to determine the effective temperature of an O-star is the \ion{He}{i} to \ion{He}{ii} ionization equilibrium \citep{1990ARA&A..28..303K}.  An investigation of the available spectral data reveals that the only \ion{He}{i} line that is clearly visible in the spectrum is at $5876$\,\AA{}. This is a result of the high temperature of both components. However, as we do not know \emph{\textup{a priori}} how much each of the binary components contributes to this line, the \ion{He}{i} to \ion{He}{ii} ionization equilibrium cannot be used to estimate the temperature. The situation is similar for other commonly used line ratios, for instance, \ion{N}{iii}/\ion{N}{iv}. Without a clear indication of the origin of the line, the line ratio is not useful. However, the absence or presence of certain lines can also give constraints on stellar parameters.

\begin{figure}
    \centering
    \includegraphics[scale=1,angle=-90]{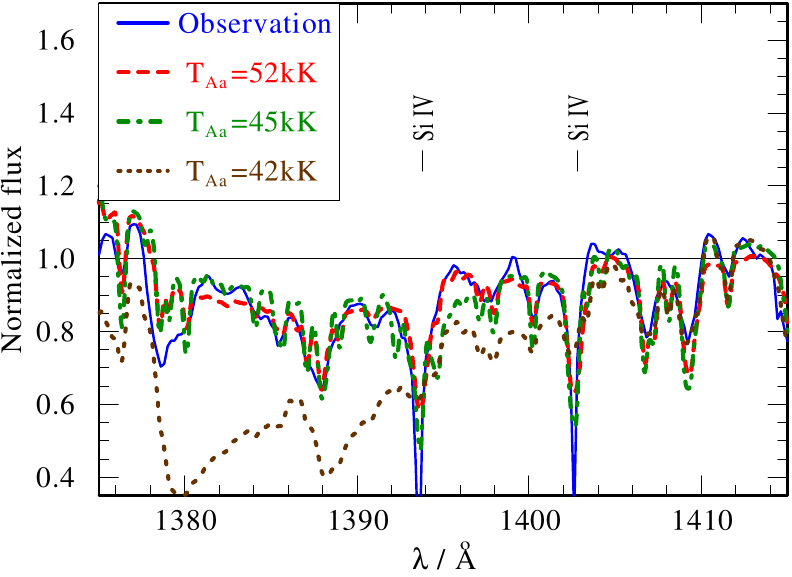} 

    \

    \includegraphics[scale=1,angle=-90]{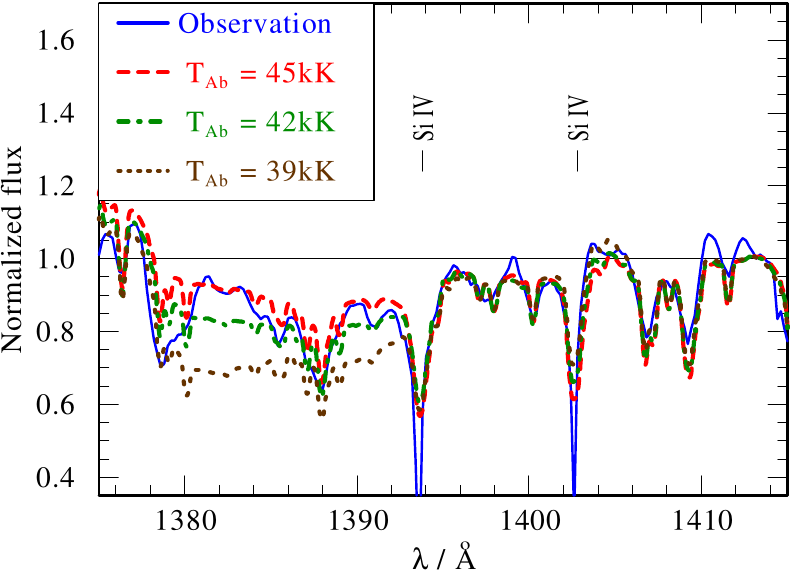} 
    \caption{Observed \ion{Si}{iv} $\lambda\lambda 1394,1403$ doublet (blue solid line) compared to synthetic spectra for Aa and Ab. In the upper panel, the temperature of the primary is varied, while the secondary's temperature is fixed at 45\,kK. In the lower panel, the secondary's temperature is varied, while the primary's temperature is fixed at 52\,kK. The red dashed lines always indicate the adopted final model. \label{fig:siv_min_temp}}
  \end{figure}

  \citet{2006IAUJD...4E..19W} presented a luminosity sequence of UV spectra for O6.5 stars (see their figure\,4). The \ion{Si}{iv} $\lambda\lambda 1394,1403$ resonance doublet is barely visible for a dwarf star, while it forms a double P-Cygni-shaped line profile for a supergiant. A comparison with \citet{1985NASRP1155.....W} shows this pattern for all subtypes of O stars, from O4 to O9. In the observation we used here, this resonance doublet is only weakly present in the wind (Fig.\,\ref{fig:siv_min_temp}). It is thus apparent that neither Aa nor Ab exhibit it strongly. To obtain synthetic \ion{Si}{iv} $\lambda\lambda 1394,1403$ profiles as observed, the models require minimum temperatures of $T_{\textrm{2/3}}>45$\,kK for Aa and of $T_{\textrm{2/3}}>41$\,kK for Ab (Fig.\,\ref{fig:siv_min_temp}). A further increase of the temperatures does not change this line significantly. Another way to obtain the weak P-Cygni profile in the models is by decreasing the mass-loss rates. However, this yields inconsistencies with other wind lines (e.g., \ion{N}{v}\,$\lambda 1238,42$, \ion{C}{iv}\,$\lambda 1548,50$, and \ion{He}{ii}\,$\lambda 4686$). Alternatively, the silicon abundance might be reduced, but this contradicts other Si features in the spectrum and lacks physical plausibility.

\begin{figure}
    \centering
    \includegraphics[scale=1,angle=-90]{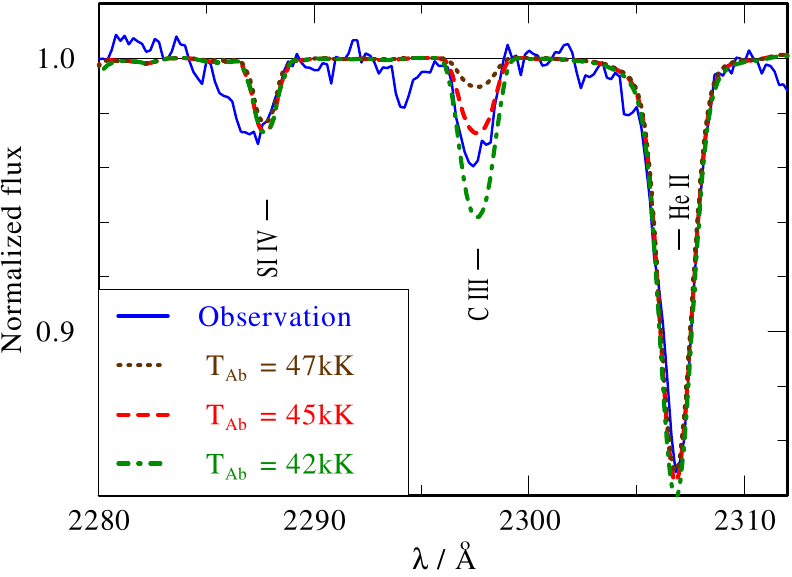} 
    \caption{Observed \ion{C}{iii} $\lambda2292$ (blue solid line) compared to models for Ab with 42\,kK (green dash-dotted line), 45\,kK (red dasehd line), and 47\,kK (brown dotted line). The  temperature of Aa is fixed at 52\,kK. \label{fig:c_iii_max_temp}}
  \end{figure}

\begin{figure}
    \centering
    \includegraphics[scale=1,angle=-90]{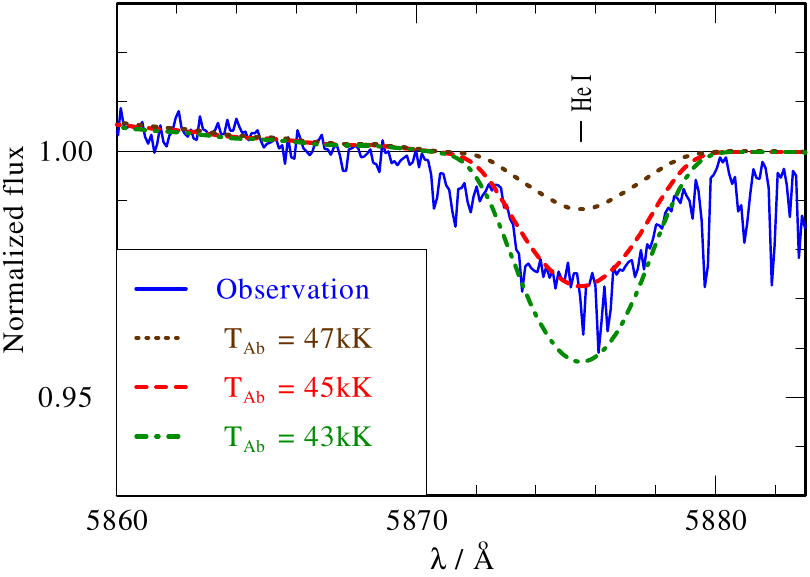}
    \caption{Observed \ion{He}{i} line $\lambda 5875$ (blue solid line) compared to models for Ab with 42\,kK (green dash-dotted line), 45\,kK (red dashed line), and 47\,kK (brown dotted line). The  temperature of Aa is fixed at 52\,kK. \label{fig:he_i_ii_line}}
  \end{figure}

\begin{figure}
    \centering
    \includegraphics[scale=1,angle=-90]{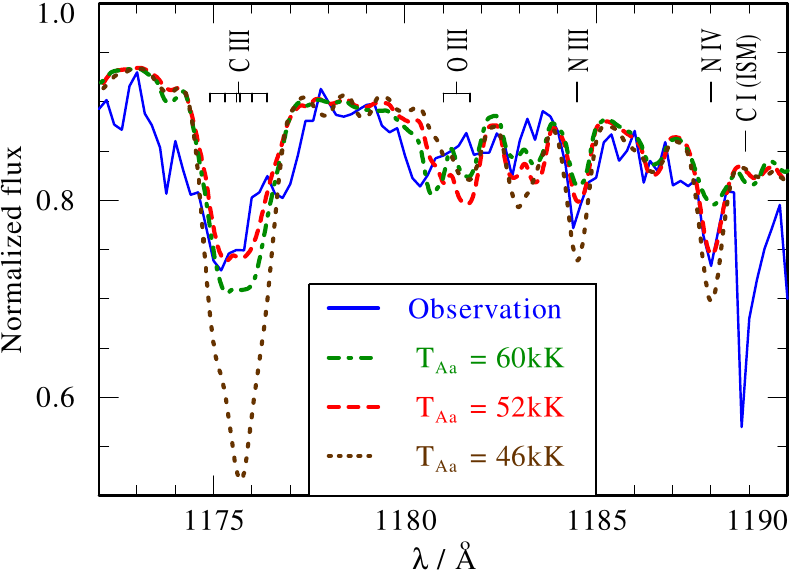}  

    \

    \includegraphics[scale=1,angle=-90]{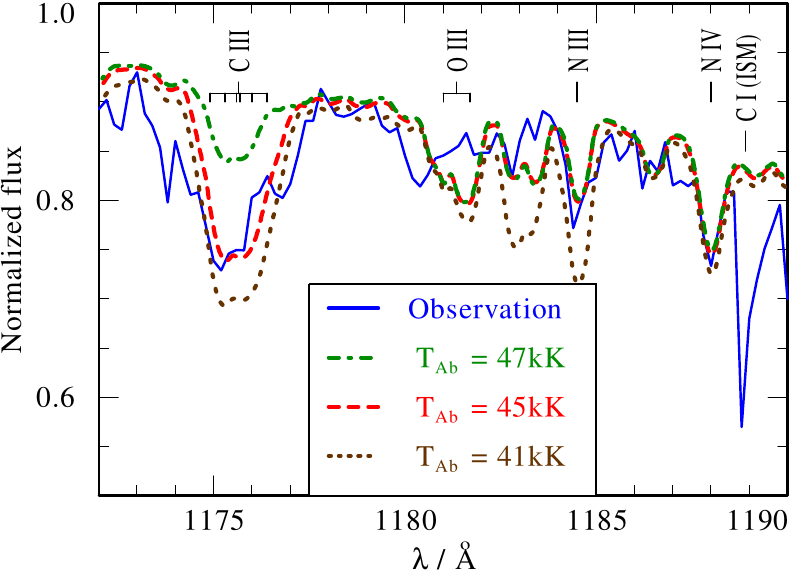}  
    \caption{Observed \ion{C}{iii} $\lambda1175$, \ion{N}{iii} $\lambda1184$, and \ion{N}{iv} $\lambda1188$ (blue solid line) compared to synthetic spectra for Aa and Ab. In the upper panel, the temperature of the primary is varied, while the secondary's temperature is kept at 45\,kK. In the lower panel, the secondary's temperatures is varied, while the primary's temperature is kept at 52\,kK. The red dashed lines indicate the final model. \label{fig:ciii_temp_1}}
  \end{figure}

  At these temperatures, our synthetic spectra predict that all \ion{He}{i}, \ion{N}{iii}, \ion{O}{iii}, and \ion{Si}{iii} lines originate only from Ab. The lines from \ion{He}{ii}, \ion{N}{iv}, \ion{C}{iii}, \ion{C}{iv}, \ion{O}{iv}, \ion{O}{v}, \ion{S}{v}, and \ion{Si}{iv} are present in both components. \ion{S}{vi} is only visible in the hotter component's spectrum. In an iterative process, where we adjusted both temperatures simultaneously in an attempt to reproduce lines of various ions along the entire spectral range, we obtained the best overall fit for $T_{2/3}^{\text{Aa}}=52$\,kK and $T_{2/3}^{\text{Ab}}=45$\,kK. We also tested scenarios with different conditions, for example, with the secondary as the hotter component or both with the same temperature. They all led to large inconsistencies for the spectral fits. Figures \ref{fig:c_iii_max_temp}-\ref{fig:ciii_temp_1} show the effect of the temperature on exemplary lines.

  Fitting the composite SED as described in Sect.\,4.1 and determining the stellar temperatures from the line spectra are in fact mutually coupled steps, which have to be iterated to consistency. We finally obtain for the bolometric luminosities of the components $\log L_{\text{Aa}}/L_\odot =6.15$ and $\log L_{\text{Ab}}/L_\odot=5.58$. The resulting fit is shown in Fig.\,\ref{fig:sed_compare}. The expected offset due to contamination between our model calculations and the observed JHK magnitudes is obvious (see Sect.\,\ref{sec:photometry}).

  As is usually found for  O-stars, HD\,93129A emits X-rays. We account for the impact of X-rays by assuming the presence of an optically thin, X-ray emitting plasma dispersed in the wind \citep{1992A&A...266..402B,2015ApJ...809..135S}. The X-ray field can in principle change the ionization in the wind. \cite{2011ApJS..194....7N} found an X-ray luminosity of log\,$(L_X/L_\odot) = -0.8$ for our target system. Adjusting the X-ray field to the observed luminosity, we find that an inclusion of X-rays in our final models has no significant impact on the synthetic spectrum.

\subsection{Rotational velocity}
 
\begin{figure}
    \centering
    \includegraphics[scale=1, angle=-90]{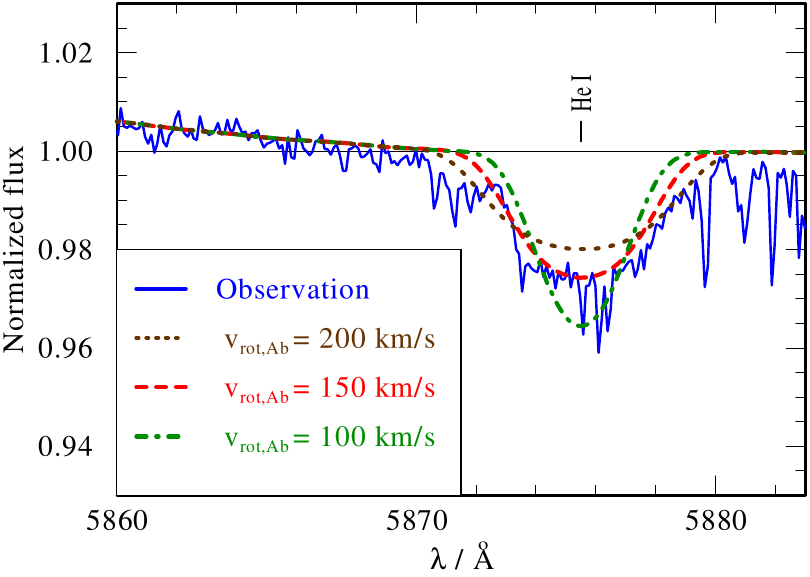} 
    \caption{Observed \ion{He}{i} $\lambda 5876$ (blue solid line) compared to composite models with rotational velocities for Ab of $v\sin i=200$\,km\,s$^{-1}$ (brown dotted line), $150$\,km\,s$^{-1}$ (red dashed line) and $100$\,km\,s$^{-1}$ (green dash-dotted line), each combined with a model for Aa with $80$\,km\,s$^{-1}$. \label{fig:rot_he_i}}
  \end{figure}

  To derive the projected stellar rotation, we compared the photospheric absorption and emission lines that originate in the spectrum of only one of the binary components with model calculations for different rotational velocities. For Ab, we chose \ion{He}{i} $\lambda 5876,$ which is the strongest line that originates from the secondary alone. Figure \ref{fig:rot_he_i} shows the absorption line profile for different rotational velocities of the secondary. We find a rotational velocity of $v_{\textrm{rot}}\sin i=150$\,km\,s$^{-1}$ for Ab. 
 
\begin{figure}
    \centering
    \includegraphics[scale=1, angle=-90]{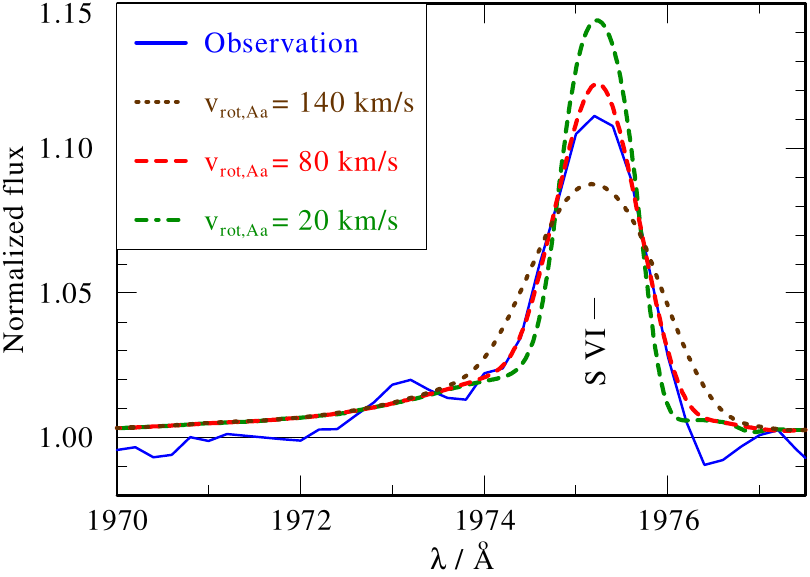} 
    \caption{Observed \ion{S}{vi} $\lambda 1975$ (blue solid line) compared to a composition of best-fitting models with rotational velocities for Aa of $v\sin i=20$\,km\,s$^{-1}$ (brown dotted line), $80$\,km\,s$^{-1}$  (red dashed line), and   $140$\,km\,s$^{-1}$ (green dash-dotted line), each combined with a model for Ab with $150$\,km\,s$^{-1}$. The red dashed spectra indicate the final model. \label{fig:rot_s_vi}}
  \end{figure}

\begin{figure}
    \centering
    \includegraphics[scale=1,angle=-90]{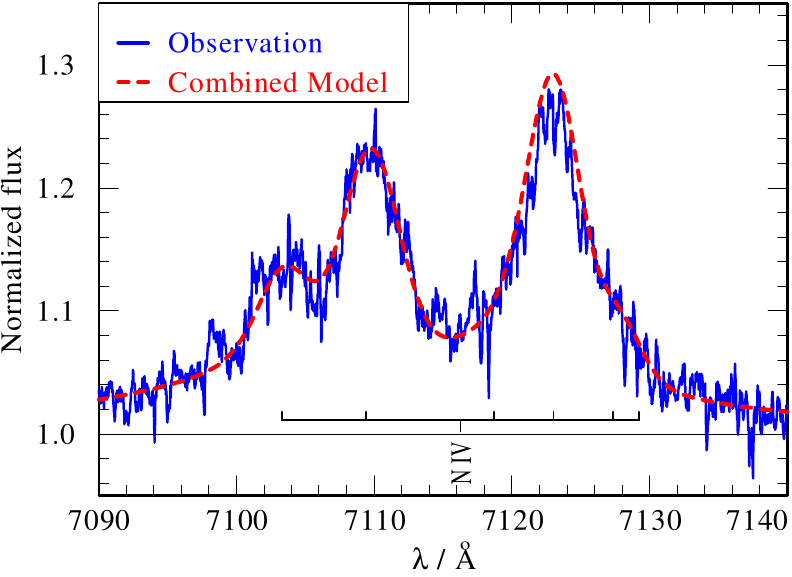} 
    \caption{Observed \ion{N}{iv} $\lambda\lambda 7103-7129 $ (blue solid line) compared to the combination of the synthetic spectra (red dashed line) for the rotational velocities of $80$\,km\,s$^{-1}$ and $150$\,km\,s$^{-1}$ for Aa and Ab, respectively. \label{fig:n_rot}}
  \end{figure}

  For Aa, we used the line \ion{S}{vi} $\lambda 1975$. Figure \ref{fig:rot_s_vi} shows the observed line profile compared to synthetic spectra with different rotational velocities of the primary. We find the best fit for $v_{\textrm{rot}}\sin i=80$\,km\,s$^{-1}$. This result is consistent with other lines such as \ion{N}{iv} $\lambda\lambda 4062,7103-7129$ and \ion{N}{v} $\lambda\lambda4603,4619$ (cf. Fig.\,\ref{fig:n_rot}).

\subsection{Abundances}

  Carbon and oxygen are found to be depleted, while nitrogen is enriched. We obtain the best agreement for an oxygen mass fraction of 0.2 solar and a carbon mass fraction of 0.05 solar for each of the components. These cannot be determined individually. 

  \begin{table}\centering
    \caption{ Overview over spectral lines used to determine the individual elemetal abundances of the components.\label{tab:abundance_lines}}
    \begin{tabular}{p{0.15\linewidth}p{.65\linewidth}}
      \hline\hline\\[-2.0ex]
      Element             & Lines \\
      \hline\\[-2.0ex]
      H                   & Balmer series, Paschen series\\
      \hline\\[-2.0ex]
      He                  & \ion{He}{i} $\lambda 5875$, Pickering series, \\
         & transitions $\#\rightarrow 3$ in \ion{He}{ii} \\
      \hline\\[-2.0ex]
      C                   & \ion{C}{iii} $\lambda\lambda 1175,$ $2297$\\
                          & \ion{C}{iv} $\lambda\lambda 1550,$ $5808 $\\
      \hline\\[-2.0ex]
      N                   & \ion{N}{iii} $\lambda\lambda 1184,$ $4634,$ $4640$,\\
                          & \ion{N}{iv} $\lambda\lambda 1188,$ $1718,$ $1721,$ $4057,$ $4609+19,$ $6214-19,$ $7103-29$\\
                          & \ion{N}{v} $\lambda\lambda 4603+19$\\
      \hline\\[-2.0ex]
      O                   & \ion{O}{iii} $\lambda 1182$\\
                          & \ion{O}{iv} $\lambda\lambda 1338-43,$ $3064+72$\\
                          & \ion{O}{v} $\lambda 1371$\\
      \hline\\[-2.0ex]
      Si                  & \ion{Si}{iii} $\lambda\lambda 2887$\\
                          & \ion{Si}{iv} $\lambda\lambda 1493,$ $1402$, $4088$, $4116$\\
      \hline\\[-2.0ex]
      S                   & \ion{S}{v} $\lambda 1501 $, \ion{S}{vi} $\lambda\lambda 1975+92$\\
      \hline\\[-2.0ex]
      Fe                  & iron forest\\
      \hline\\[-2.0ex]
      P                   & no lines in the spectral range\\
      \hline
    \end{tabular}
  \end{table}
 
\begin{figure}
    \centering
    \includegraphics[scale=1,angle=-90]{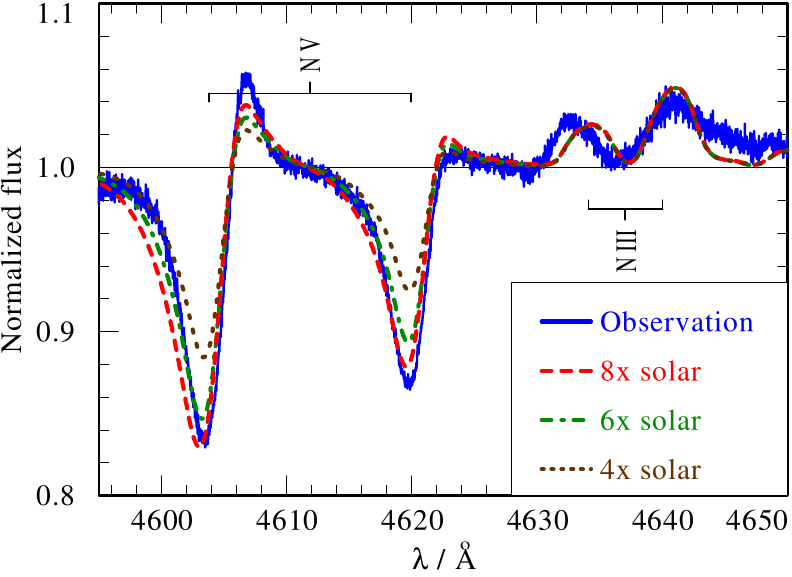} 

    \

    \includegraphics[scale=1,angle=-90]{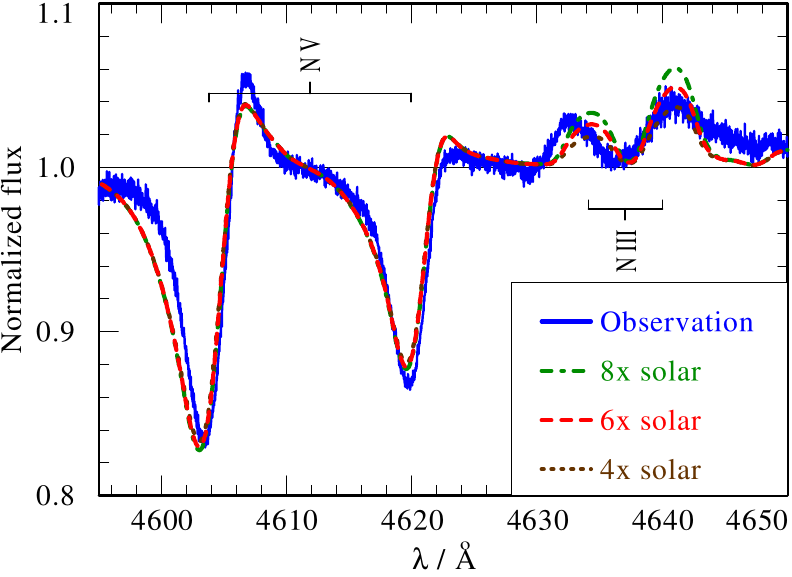} 
    \caption{Observed \ion{N}{v} $\lambda\lambda4603,4619$ and \ion{N}{iii} $\lambda\lambda4634,4640$ (blue solid line) compared to best-fitting model composition. In the upper panel, the nitrogen abundance of the primary is varied, while the secondary's nitrogen abundance is kept at six times solar. In the lower panel, the secondary's nitrogen abundance is varied, while the primary's nitrogen abundance is kept at eight times solar. The red dashed lines indicate the final model.  \label{fig:abu_n_iv}}
  \end{figure}

  We are able to determine the abundances of nitrogen for each component separately, owing to the three different ionization stages present in the spectrum. We derive a nitrogen abundance that is eight times solar for Aa and six times solar for Ab (Fig.\,\ref{fig:abu_n_iv}). These results are consistent with the fact that the primary in the system is classified as a supergiant and is therefore expected to be chemically evolved. We do not detect a significant He enrichment from the spectra. For sulfur, silicon, and the iron group elements, we assume solar abundances and achieve a good agreement. The observed spectrum shows no significant phosphorus lines.  We therefore assume solar abundance for phosphorus as well. The effect of iron on the spectra can be seen in an extend region in the UV, where the spectrum shows numerous iron lines, and which is called the \emph{\textup{iron forest}}.

  An overview of the lines used for the abundance determination of the chemical elements is given in Table\,\ref{tab:abundance_lines}.

\subsection{Mass-loss and clumping}

  We attempted to derive the clumping factor $D$ for the primary star by comparing the strengths of the absorption part of the resonance P-Cygni profiles, which scale with $\dot{M}$, and recombination emission lines, which scale as $\dot{M} \sqrt{D}$. However, there are no unsaturated resonance lines in the HST spectrum. The unsaturated P-Cygni lines that are visible in the spectrum are not resonance transitions, and therefore change with $D$ and $\dot{M}$ in a complex way.

\begin{figure}
    \centering
    \includegraphics[width=\linewidth]{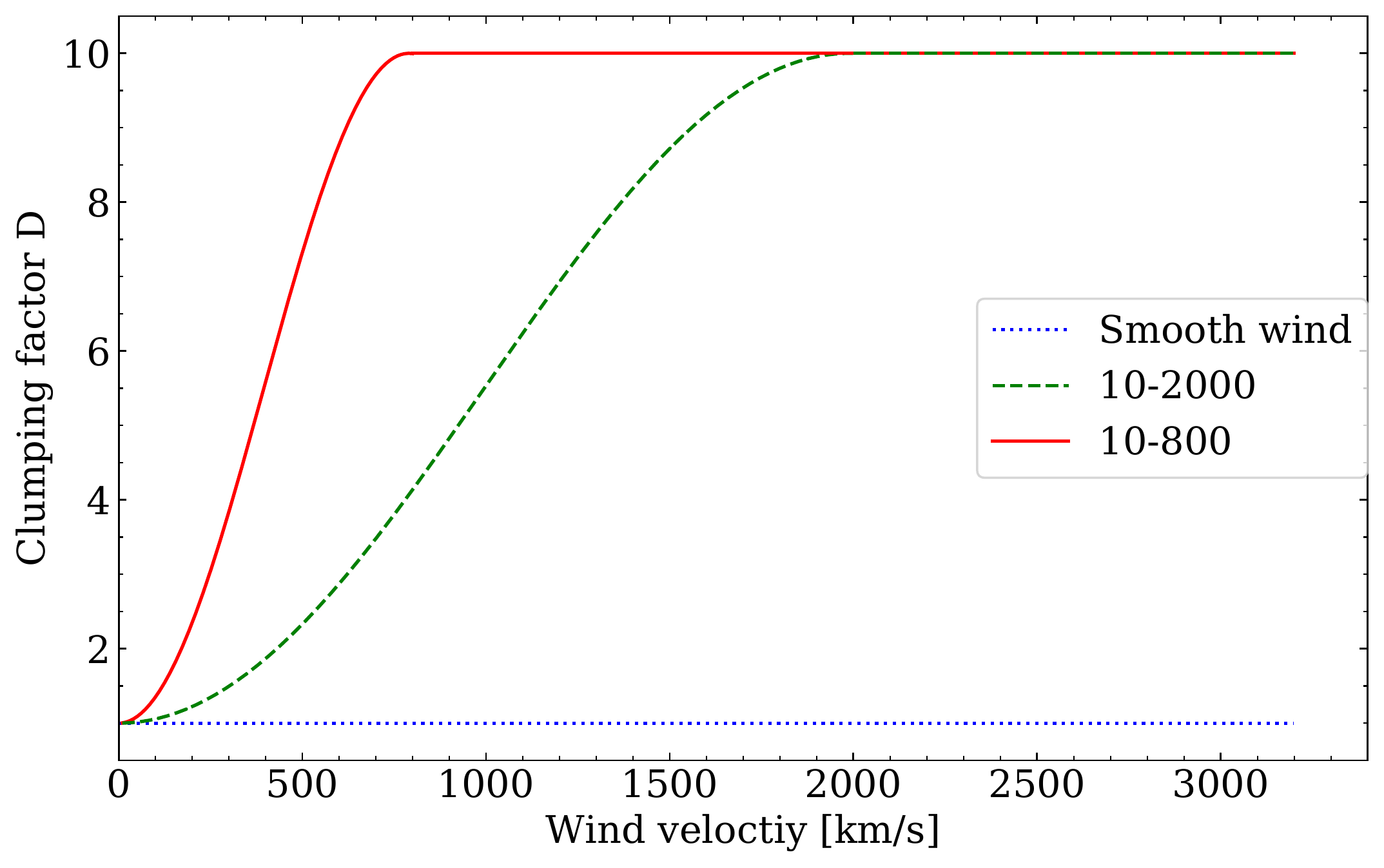} 
    \caption{ Adopted clumping stratification for the models. (10--800) and (10--2000) denote the velocity range in which the clumping is set to grow from a smooth wind ($D=1$) to the maximum value (here $D=10$). After reaching its maximum value, $D$ is kept constant. \label{fig:clumbs}}
  \end{figure}

\begin{figure}
    \centering
    \includegraphics[scale=1,angle=-90]{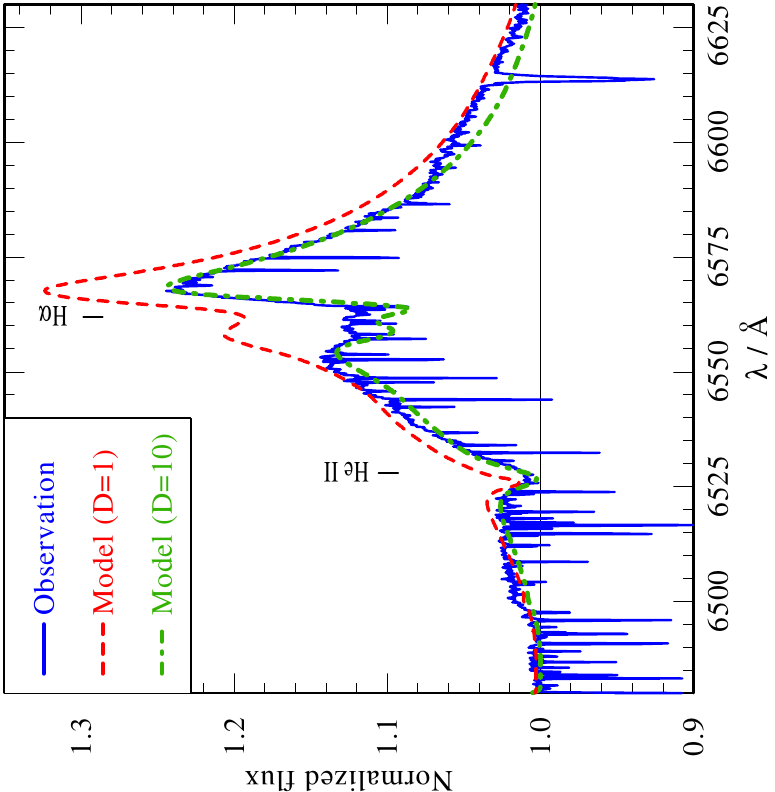} 
    \caption{ Observed H$\alpha$  (blue solid line) compared to the final composite model with $D=1$ for the primary (dashed red line)  and a composite model with radial dependent $D$ as displayed by the red dashed line in Fig.\,\ref{fig:clumbs} (dash-dotted green line).\label{fig:massloss}}
  \end{figure}

  When adopting a clumping stratification as described in Sect.\,\ref{sec:powr} (cf. the red line in Fig\,\ref{fig:clumbs}), the spectrum is reproduced well, while H$\alpha$ is reproduced poorly. A reduction of the clumping factor leads to an increased fit quality for H$\alpha$, but conserves the fit quality in the overall spectrum. The best overall fit is achieved when no clumping at all is assumed for Aa (cf. Fig.\,\ref{fig:massloss}). However, this seems to contradict various studies that generally find evidence for clumping factors of $D \gtrsim 10$ in OB-type stars \citep[e.g.,][]{2012A&A...544A..67B}.

  We were able to achieve a good fit for H$\alpha$ for $D=10$ (green line in Fig.\,\ref{fig:massloss}), where $D$ is assumed to increase with the wind velocity from $D=1$ at the photosphere and reached maximum value at wind velocities  $>2000\,{\rm km\, s}^{-1}$ (cf. the green line in Fig.\,\ref{fig:clumbs}). When this clumping stratification is adopted, the fit quality over the entire spectral range decreases significantly.   This may indicate that the assumptions made here for the clumpiness of the wind are overly simplified.

  Because of this uncertainty of $D$, we prefer to give only $\dot{M} \sqrt{D}$  as our results in Table\,\ref{tab_main_results}. Thus, the given value is an upper limit to the mass-loss rate, while the actual $\dot{M}$ may be a factor 2--3 lower, depending on the true clumping factor.

  \begin{figure}
    \centering
    \includegraphics[scale=1,angle=-90]{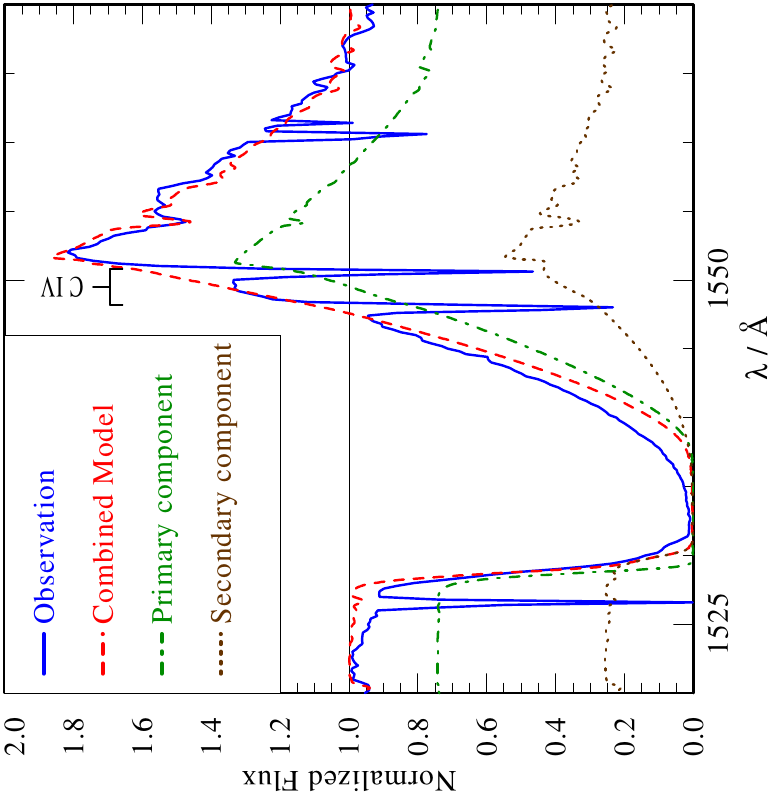} 

    \

    \includegraphics[scale=1,angle=-90]{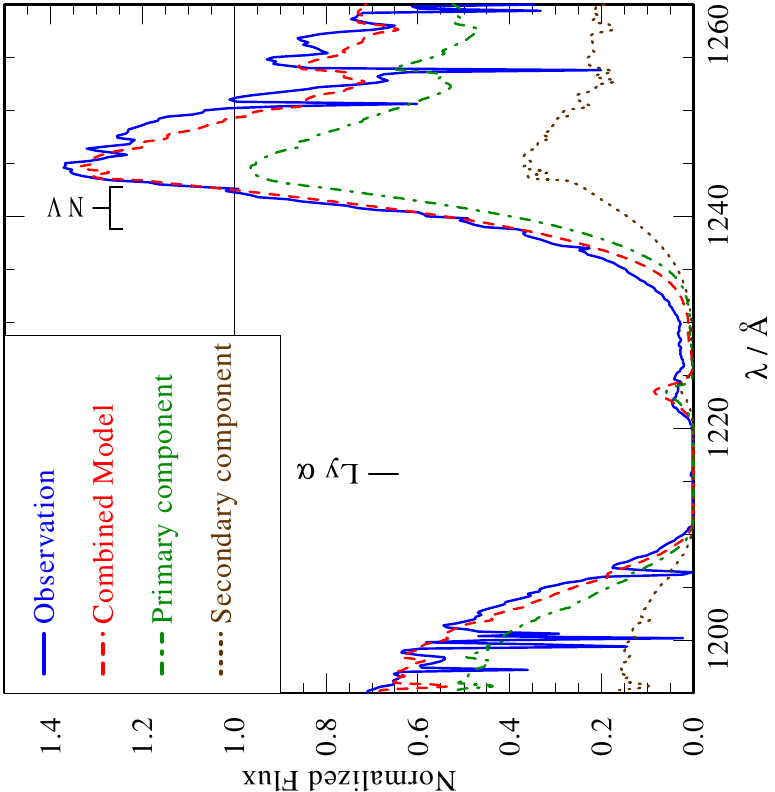}  
    \caption{ Resonance line doublets: observed  \ion{C}{iv} $\lambda 1550$ (upper panel) and \ion{N}{V} $\lambda 1245$ (lower panel) (blue solid line) compared to the best-fitting model (red dashed line). The green dash-dotted line and the brown dotted line show the flux-weighted contribution of the components Aa and Ab, respectively. \label{fig:windlines}}
  \end{figure}

\subsection{Wind velocity field}

  The terminal velocity of a stellar wind is usually inferred from the blue edge of the absorption part of P-Cygni line profiles, while the microturbulence defines the position of the emission peak and the steepness of the blue edge. For a binary star, this is again more complicated.  In principle, a saturated P-Cygni line in a binary can only form if both stars exhibit saturated lines in their intrinsic spectra. If a difference in $v_\infty$ exists, this difference is manifested through a ``step'' in the composite P-Cygni profile \citep[see, e.g., figure 5 in][]{2016A&A...591A..22S}. Such steps are not observed here. We therefore conclude that the winds of Aa and Ab have very similar (within 100\,\kms)  terminal velocities.  Similarly, there is no way to distinguish between the microtubulence $v_\text{mic}$ of stars Aa and Ab.

  Testing different values for the $\beta$ parameter of the velocity law, we find the best agreement with observed lines for $\beta = 1.2$ and $v_\infty = 3200\,$\kms. This result is compatible with previous reports for the object \citep{1997A&A...321..531T}. The microturbulence is assumed to grow up to 300\,\kms (approx.\ 10\% of the terminal velocity). The same $\beta$ and $v_\text{mic}$ are adopted for both components.

  For the radial velocity, we applied in  principle an individual wavelength shift for the components before adding the spectra. Through fitting line peaks and absorption profiles, we determined the radial velocity separately for both stars. However, we obtained the same velocity for both stars: $v_{\text{rad}}=5\pm10\,$km\,s$^{-1}$, consistently with the results reported by \citet{2017MNRAS.464.3561M}.

\section{Results} \label{sec:results}

\begin{figure*}
    \centering
    \includegraphics[scale=1,angle=-90]{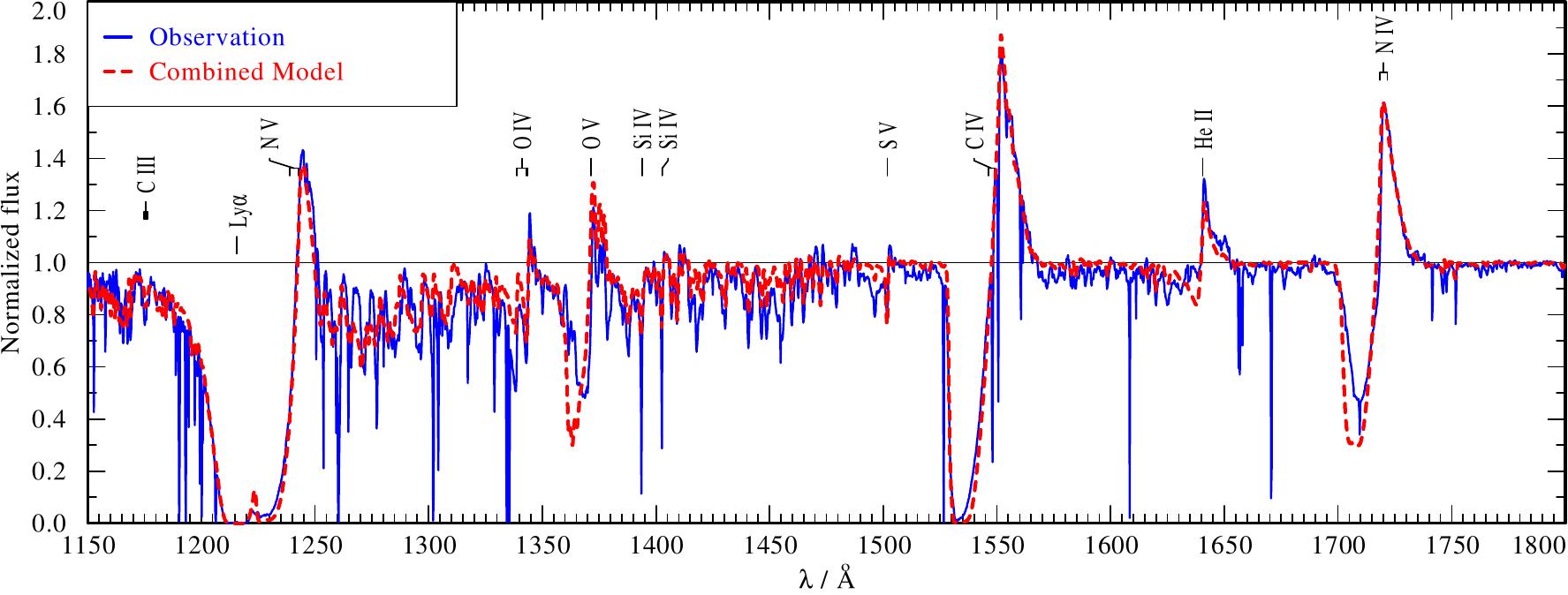} \\
    \includegraphics[scale=1,angle=-90]{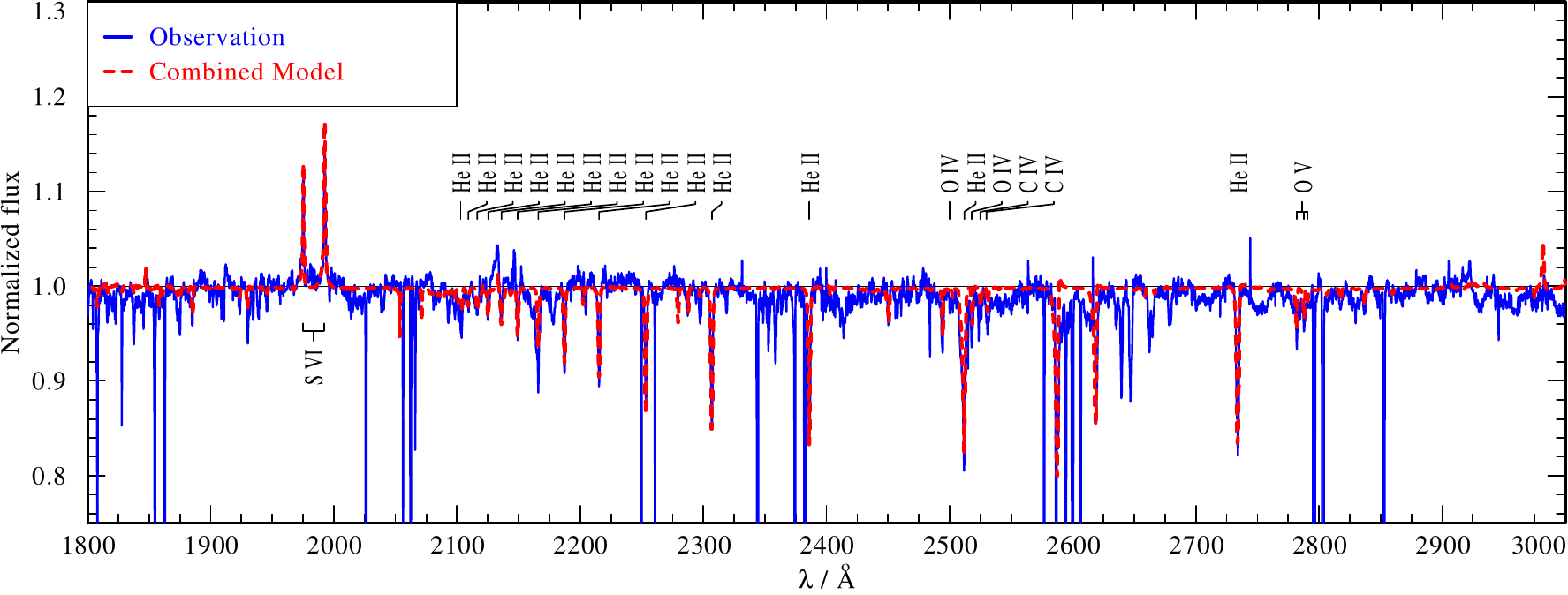} 
    \caption{HST observation of HD\,93129A (blue solid line) compared to our best-fitting composed model (red dashed line). The HST data were binned to 0.1\,\AA{} for clarity. \label{pic:complete_uv}}
  \end{figure*}

\begin{figure*}
    \centering
    \includegraphics[scale=1,angle=-90]{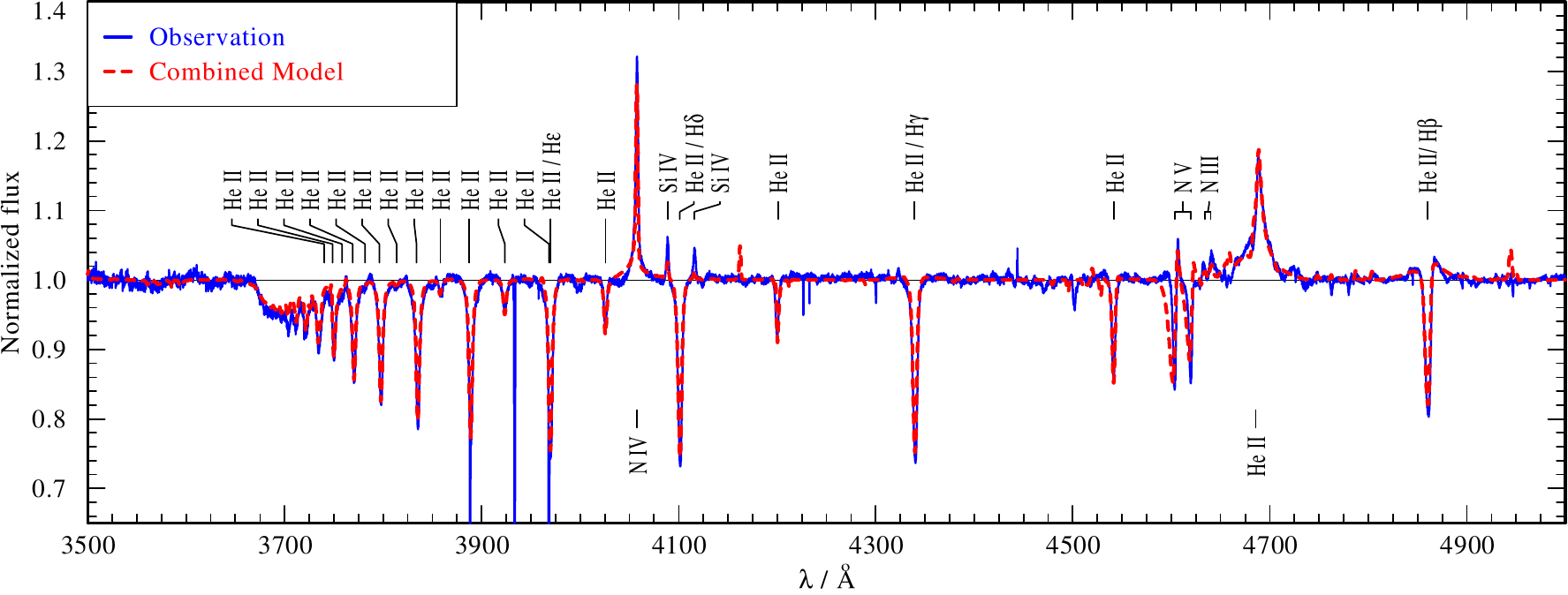} \\
    \includegraphics[scale=1,angle=-90]{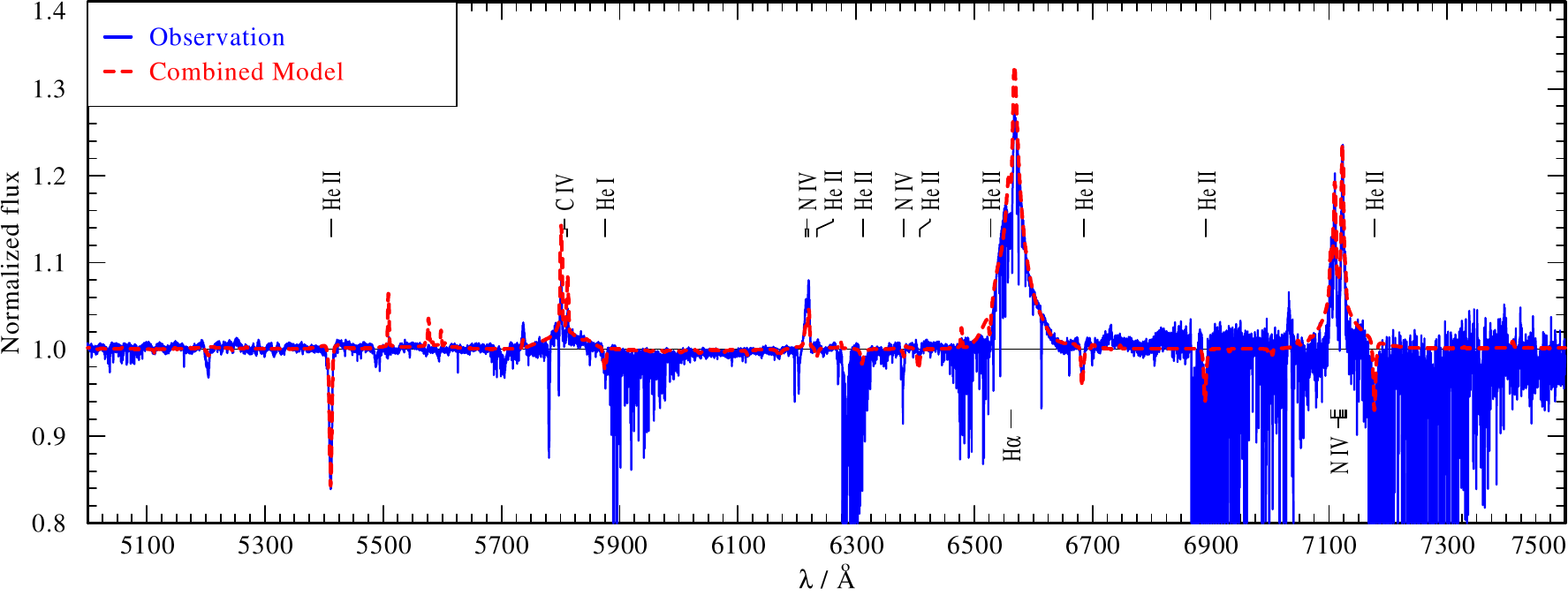} \\
    \includegraphics[scale=1,angle=-90]{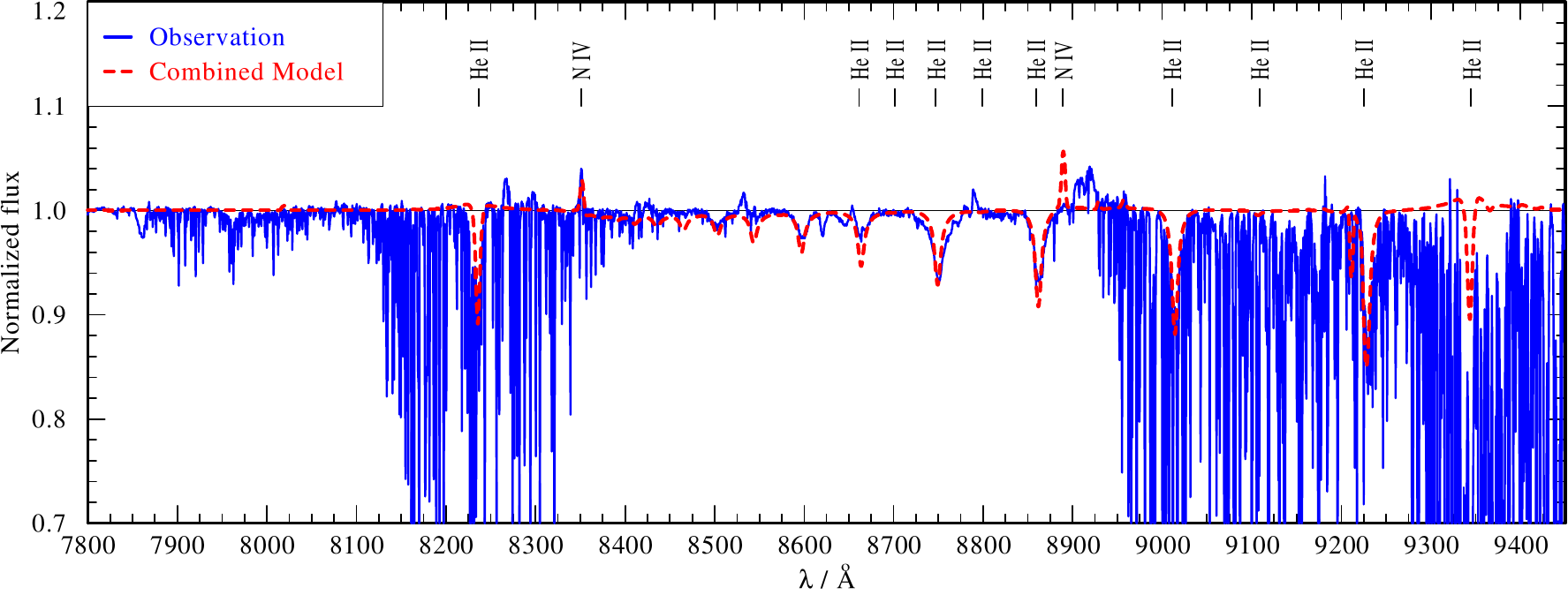} 
    \caption{UVES observation of HD\,93129A (blue solid line) compared to our best-fitting composite model (red dashed line).\label{pic:complete_opt}}
  \end{figure*}

  Table \ref{tab_main_results} summarizes the parameter of the best-fitting models for Aa and Ab. The stellar radius $R_*$ was calculated from the Stefan-Boltzmann law, while the spectroscopic mass $M_*$ follows from the radius and surface gravity. Table \ref{tab_abund_results} lists the abundances, and Table \ref{tab_photo_results} gives the resulting absolute magnitudes from our spectral synthesis for the components. Figures \ref{pic:complete_uv} and \ref{pic:complete_opt} show the normalized observation compared to the best fit, combining the synthetic spectra of the primary and secondary.

  \begin{table}
    \centering
    \caption{Parameters of the best-fitting models for Aa and Ab. \label{tab_main_results}}
    \begin{tabular}{lcll}
      \hline\hline
      Star   \rule[0mm]{0mm}{4mm}     &                  & Aa                               & Ab \\
      \hline
      $T_{\textrm{2/3}}$ \rule[0mm]{0mm}{5mm}  & [kK]          & $52.0\pm3.0$              &   $45.0\pm3.0$         \\
      log\,$L$             & [$L_\odot$]        & $6.15\pm0.05$                    &   $5.58\pm0.05$         \\
      $M_*$  & [$M_\odot$]                              & $100\substack{+25\\-60}$                       &    $37\substack{+15\\-20}$           \\
      $R_*$ &  [$R_\odot$]                              & $14.7\pm2$                       &    $10\pm2$          \\
      log\,$g$               &          [cm\,s$^{-2}$]  & $4.1\substack{+0.1\\-0.4}$                      &   $4.0\substack{+0.2\\-0.2}$          \\
      log\,$(\dot M\sqrt{D})$ &  [$M_\odot$\,yr$^{-1}$]  & $-4.7\pm0.05$           &   $-5.8\pm0.2$              \\
      $v_\infty$ & [km\,s$^{-1}$]               & $3200\pm100$                     &   $3200\pm100$             \\
      $\beta$              &                    & $1.2$                            &   $1.2$                 \\
      $v_{\textrm{rot}}\sin i$  & [km\,s$^{-1}$]                & $80\pm10$  &   $150\pm10$   \\ 
      $v_{\textrm{rad}}$   & [km\,s$^{-1}$]                     & $5\pm10$    &    $5\pm10$             \\
      \hline
    \end{tabular}
  \end{table}

  \begin{table}
    \centering
    \caption{Chemical abundances for Aa and Ab. The phosphorus abundance could not be derived in our analysis and is assumed to be solar.  Error estimates correspond to upper and lower limits where the change in abundance starts to become noticeable in the spectrum \label{tab_abund_results}}
    \begin{tabular}{ccc}
      \hline\hline
      \rule[0mm]{0mm}{4mm}
      Element             & \multicolumn{2}{c}{Mass fraction}  \\
                          &        Aa       & Ab \\
      \hline
      \rule[0mm]{0mm}{4mm}
      H  & $(7.3\pm 0.3)\cdot 10^{-1}$ & $(7.3\pm 0.3)\cdot 10^{-1}$ \\
      He & $(2.5\pm 0.3)\cdot 10^{-1}$ & $(2.5\pm 0.3)\cdot 10^{-1}$ \\
      C  & $(1.2\pm 0.2)\cdot 10^{-4}$ & $(1.2\pm 0.2)\cdot 10^{-4}$ \\
      N  & $(5.6\pm 0.4)\cdot 10^{-3}$ & $(4.2\pm 0.4)\cdot 10^{-3}$ \\
      O  & $(1.1\pm 0.5)\cdot 10^{-3}$ & $(1.1\pm 0.5)\cdot 10^{-3}$ \\
      Si & $6.6\cdot 10^{-4}$ & $6.6\cdot 10^{-4}$ \\
      (P)& $5.8\cdot 10^{-6}$ & $5.8\cdot 10^{-6}$ \\
      S  & $3.1\cdot 10^{-4}$ & $3.1\cdot 10^{-4}$ \\
      Fe & $1.4\cdot 10^{-3}$ & $1.4\cdot 10^{-3}$ \\
      \hline
    \end{tabular}
  \end{table}

  \begin{table}
    \centering
    \caption{Absolute magnitudes from calculations for the best-fitting combination of models   \label{tab_photo_results}}
    \begin{tabular}{lcccc}
      \hline\hline
      Band   \rule[0mm]{0mm}{4mm}               & Aa            & Ab               & Aa+Ab     & $\Delta m$ \\
                                                & [mag]         & [mag]            & [mag]     & [mag]        \\
      \hline
      U   \rule[0mm]{0mm}{4mm}                  & $-7.62$       &   $-6.71$       &  $-8.01$  & $-0.91$ \\
      B                                         & $-6.40$       &   $-5.52$       &  $-6.80$  & $-0.88$ \\
      V                                         & $-6.09$       &   $-5.21$       &  $-6.49$  & $-0.88$ \\
      J                                         & $-5.44$       &   $-4.46$       &  $-5.81$  & $-0.98$ \\
      H                                         & $-5.45$       &   $-4.40$       &  $-5.80$  & $-1.05$ \\
      K                                         & $-5.41$       &   $-4.27$           &  $-5.74$  & $-1.14$ \\
      \hline
    \end{tabular}
  \end{table}

\subsection{Spectral classification}
     
  With the decomposed spectra, we are able to reconsider the spectral classification.  The preliminary classification by \citet{2002AJ....123.2754W} is based on the composite spectra and on the extreme luminosity associated with HD\,93129A. After the discovery of the binarity of HD\,93129A, the class O2~I was assigned to the primary Aa \citep{2004AJ....128..323N}. We can now take advantage of our disentangled synthetic spectra to perform the classification, following classification schemes by  \citet{1990PASP..102..379W}, \citet{2002AJ....123.2754W}, \citet{2006IAUJD...4E..19W}, and \citet{2011ApJS..193...24S}.
 
\begin{figure*}\centering
    \includegraphics[angle=-90,width=\linewidth]{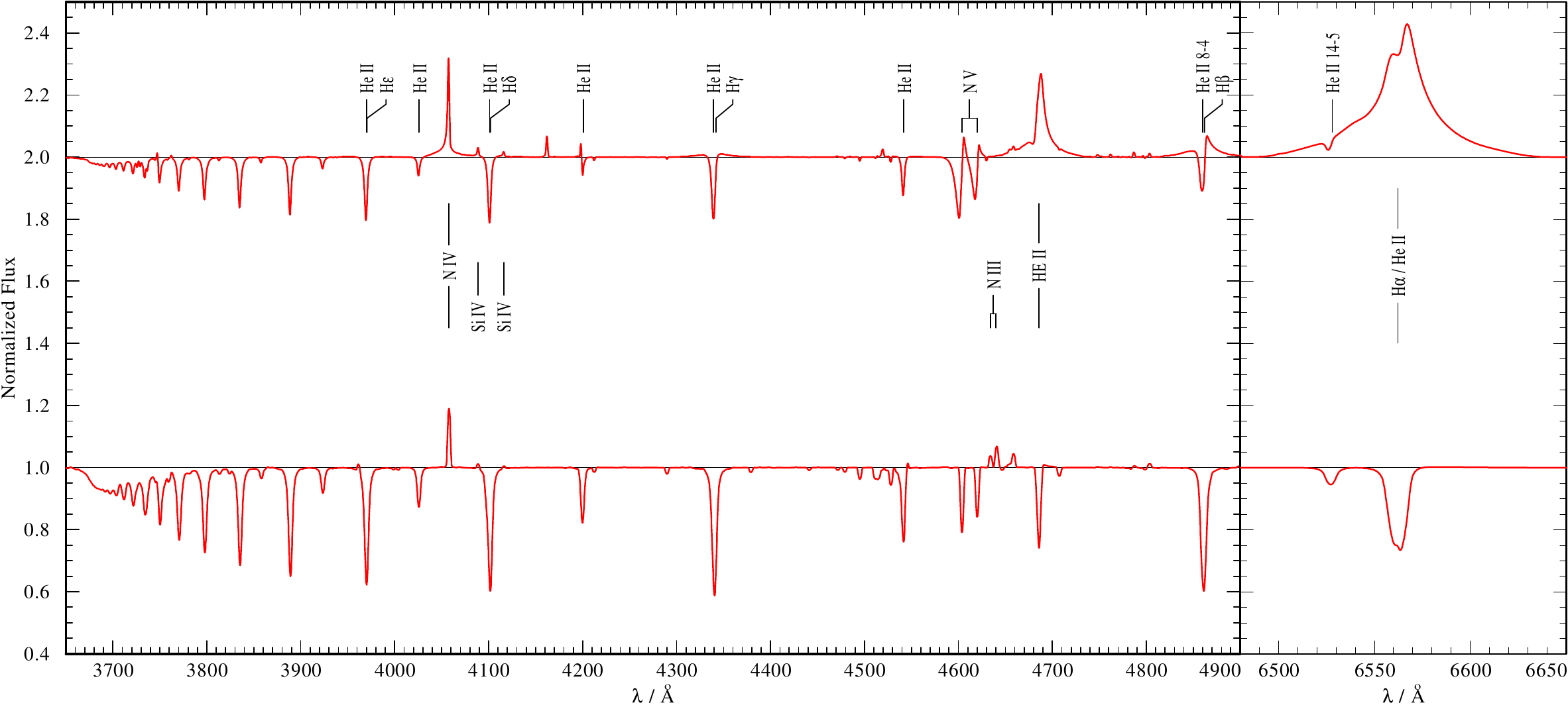} 
    \caption{Normalized, synthetic disentangled spectra for HD\,93129Aa and Ab.  For clarity, the spectrum of Aa is shifted by one flux unit. \label{fig:model_spec}}
  \end{figure*}

  Figure \ref{fig:model_spec} shows the optical range that is commonly used for spectral classification. The spectrum of Aa exhibits strong \ion{N}{v} $\lambda\lambda 4603,4619$ and  \ion{N}{iv} $\lambda\lambda 4057,7103-7129$ emission, but hardly any \ion{N}{iii} $\lambda\lambda 4634,4640$ emission, while the H$\beta$ absorption line is partly filled by emission.  Therefore, Aa fulfills the classification criteria of O2~If* or O2~If*/WN5 \citep[see, e.g., \object{Mk~42}  and \object{Mk~35} in figures 1 and 3 of][]{2011MNRAS.416.1311C}. Since the H$\beta$ emissions is weak, we classify Aa as O2~If*.

  In the spectrum of Ab we find the following characteristics: the \ion{He}{ii} $\lambda 4686$ line shows a weak emission with small self-absorption, which indicates a giant star.  The \ion{N}{iv} emission is stronger than the \ion{N}{iii} emissions by a factor of 2, \ion{N}{v} $\lambda\lambda 4603,4619$ is in absorption. A comparison to \cite{2002AJ....123.2754W} reveals the closest similarity to \object{Cyg OB2-22AB,} and Ab is therefore classified as an O3 III(f*) star.

\subsection{Stellar feedback}

  The stellar feedback of this exceptional binary is enormous (see Table \ref{tab:feedback}). A comparison to the census for the stellar content of Tr\,14 made by  \citet{2006MNRAS.367..763S} reveals that the primary alone contributes about the half of the hydrogen ionizing photons of the cluster.

  \begin{table}[ht!]\centering
    \caption{Number of ionizing photons emitted by HD\,93129A \label{tab:feedback}}
    \begin{tabular}{lcccc}
      \hline
      \hline
      \rule[0mm]{0mm}{4mm}                      & Ionization  & log$(Q_\text{Aa})$ & log$(Q_\text{Ab})$ & log$(Q_{\text{total}})$ \\
                                                & Edge [\AA]         &     [$1/$s]   &    [$1/$s] &    [$1/$s] \\
      \hline
      \ion{H}{i} \rule[0mm]{0mm}{4mm}           & 911.5 & 49.96 & 49.29 & 50.04 \\
      \ion{He}{i}                               & 504.3 & 49.45 & 48.50 & 49.50 \\
      \ion{He}{ii}                              & 227.8 & 41.85 & 39.96 & 41.86 \\
      \ion{O}{ii}                               & 353.0 & 48.71 & 47.46 & 48.73 \\
      \hline
    \end{tabular}
  \end{table}

  Furthermore, there is a large amount of mechanical energy transferred by the stellar wind. For Aa we derive a mechanical luminosity
  \begin{equation*}
    P_{\text{mech}}=\frac{1}{2}\dot M v_{\infty}^2=6.48\cdot 10^{37}\text{erg\,s}^{-1}
  \end{equation*}
  and $P_{\text{mec}}=5.14\cdot 10^{36}\text{erg\,s}^{-1}$ for Ab. The mechanical power of the binary system thus corresponds to $10 ^{4.26} L_\odot$.

\section{Discussion} \label{sec:discussion}

  \begin{figure*}
    \centering
    \includegraphics[scale=1]{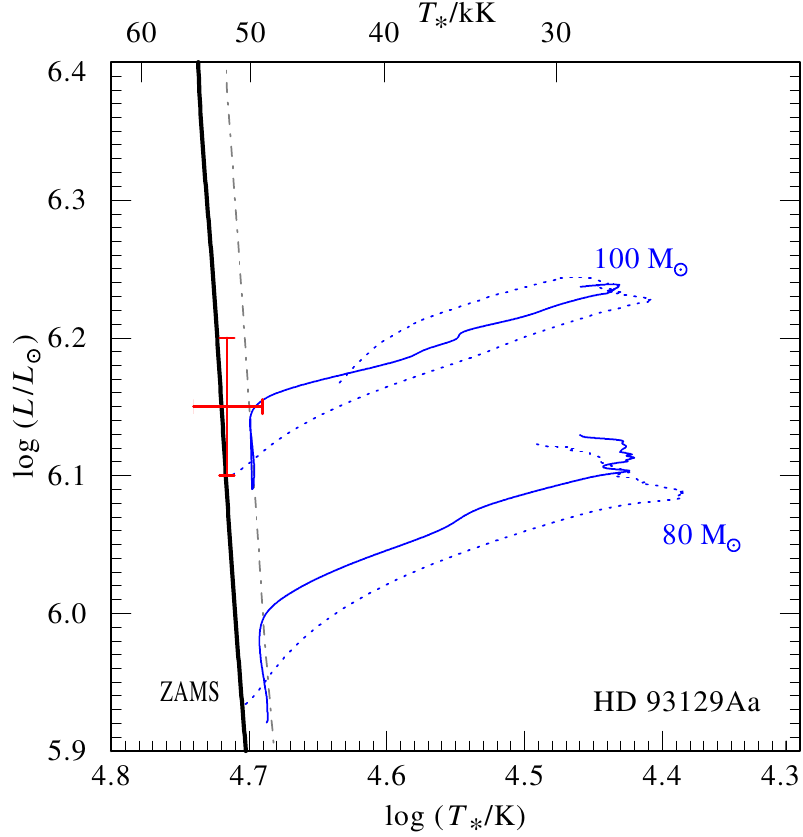} 
    \hspace{2em}
    \includegraphics[scale=1]{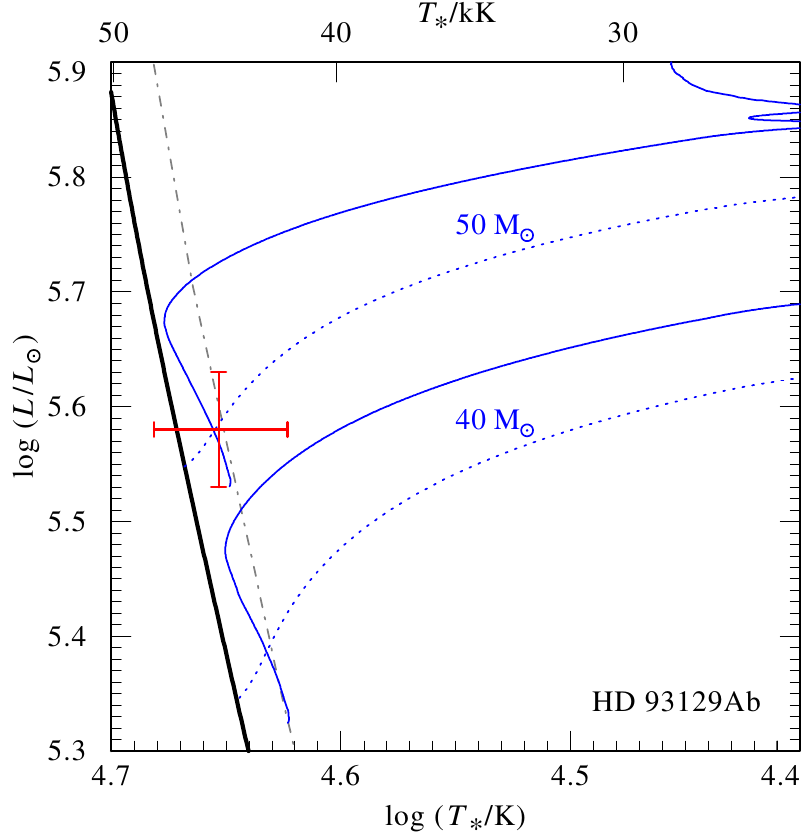} 

    \

    \includegraphics[scale=1]{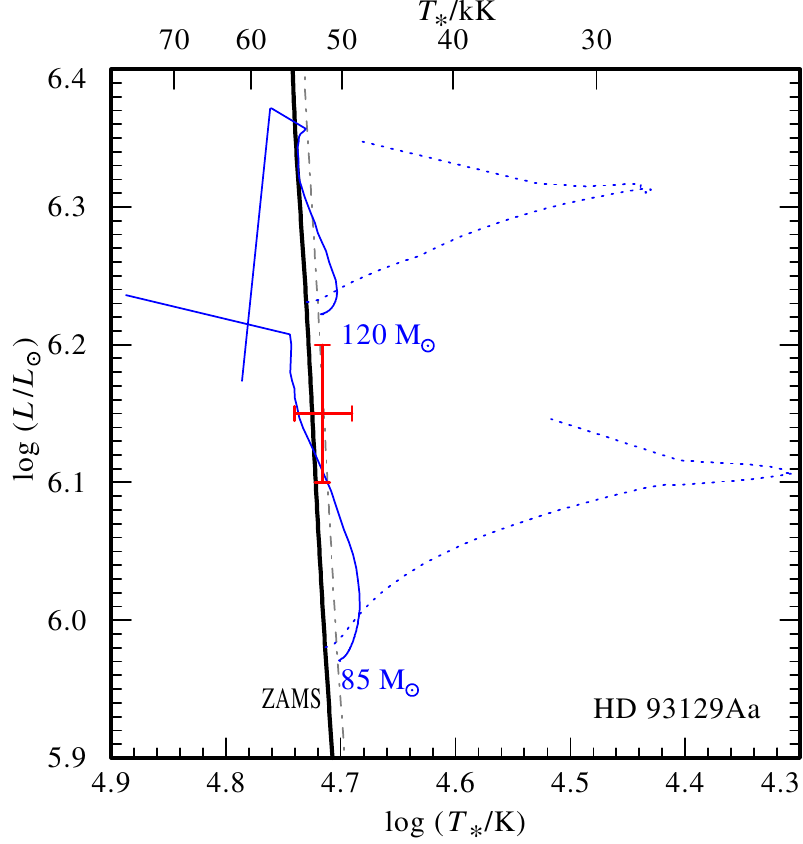} 
    \hspace{2em}
    \includegraphics[scale=1]{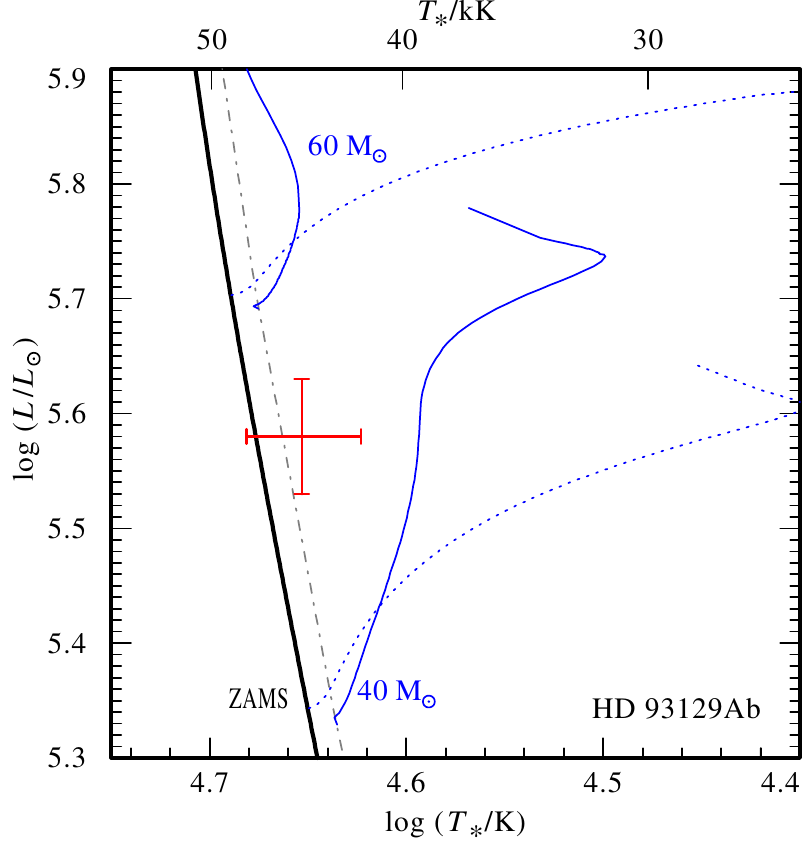} 
    \caption{HRDs for Aa (left) and Ab (right) compared to the ZAMS (solid black and gray dashed lines for non-rotating and rotating stars, respectively) and evolutionary tracks calculated by \citet{2011A&A...530A.115B} (upper panels) and \citet{2012A&A...537A.146E} (lower panel), respectively. The blue solid lines depict the tracks for non-rotating stars, and the blue dotted lines show tracks for rotating stars. All tracks show the pre-He burning phase. The positions of the components of HD\,93129A from our analysis are shown in red. \label{fig:hrd} }
  \end{figure*}
 
  To place our results in an evolutionary context, we compared the derived $T_*,L,$ and $M_*$ to tracks resulting from stellar evolution models. Given the large spatial separation of the two components, we neglected possible binary interaction in the system. In Fig.\,\ref{fig:hrd}, Hertzsprung-Russell diagrams (HRD) are shown that depict different sets of tracks, with rotationally induced mixing and without, as calculated by \citet{2011A&A...530A.115B} and \citet{2012A&A...537A.146E}. Tracks for rotating stars were calculated with initial $v_\textrm{rot}=450\,$km\,s$^{-1}$ \citep{2011A&A...530A.115B}, while \citet{2012A&A...537A.146E} adopted 40\% of the critical velocity, which describes the rotation when the gravitational acceleration is exactly counterbalanced by the centrifugal force.
 
  The temperature and luminosity obtained from our analysis indicate for both stars a position close to the main sequence.  According to the tracks, there are two possibilities for each of the components: either the star is very young and massive, or it is of lower mass, but in an evolved state, heading toward the Wolf-Rayet phase. However, we can eliminate the second possibility based on the derived H and He abundances. We thus conclude that both stars are very young and massive. 

  To derive the best-fitting evolutionary parameters for the two components, we applied the BONNSAI\footnote{The BONNSAI web-service is available at \url{www.astro.uni-bonn.de/stars/bonnsai.}} tool \citep{2014A&A...570A..66S}. The best-fitting evolutionary parameters are listed in Table\,\ref{tab_brott_parameter}.  Our results fall within the error margins reported by the BONNSAI tool.

  \begin{table}
    \centering
    \caption{Model predictions by evolutionary tracks from \citet{2011A&A...530A.115B}  with BONNSAI that fit best to our results. The underlying evolutionary tracks are limited to an initial mass $\leq100\,M_\odot$ \label{tab_brott_parameter} }
    \begin{tabular}{llll}
      \hline\hline
      \rule[0mm]{0mm}{5mm}    &                & Aa         & Ab \\
      \hline
      $T_{\textrm{eff}}$ & [kK] \rule[0mm]{0mm}{5mm}   & $49.6$     &   $44.8$         \\
      log\,$L$   &[$L_\odot$]                          & $6.10$     &   $5.58$         \\
      $M_{*,\text{act}}$  &[$M_\odot$]                   & $95.0$     &   $46.8$           \\
      $M_{*,\text{ini}}$  &[$M_\odot$]                   & $99.8$     &   $48.4$           \\
      $R_*$       &[$R_\odot$]                   & $15.4$     &   $10.2$          \\
      log\,$g$      &  [cgs]                             & $4.05$     &   $4.1$          \\
      $v_{\textrm{rot,act}}$ & [$\textrm{km\,s}^{-1}$] & $80$       &   $160$   \\ 
      $v_{\textrm{rot,ini}}$ & [$\textrm{km\,s}^{-1}$] & $90$       &   $170$   \\ 
      $X_\text{He}$& mass fraction & $2.66\cdot 10^{-1}$ & $2.66\cdot 10^{-1}$ \\
      $X_\text{C}$ & mass fraction & $1.18\cdot 10^{-3}$ & $1.18\cdot 10^{-3}$ \\
      $X_\text{N}$ & mass fraction & $4.56\cdot 10^{-4}$ & $4.56\cdot 10^{-4}$ \\
      $X_\text{O}$ & mass fraction & $4.14\cdot 10^{-3}$ & $4.18\cdot 10^{-3}$ \\
      \hline
    \end{tabular}
  \end{table}

  Comparing these values with our spectroscopic results (Table\,\ref{tab_main_results}),  we see that good agreement is obtained for the primary. The slight difference in parameters can be attributed to the fact that the highest initial mass considered by \citet{2011A&A...530A.115B} is 100$\,M_\odot$. Their results for luminosity, temperature, and mass are therefore slightly lower than obtained here. Judging by the primary's derived age and its associated error  of $0.20\substack{+0.51\\-0.19}$\,Myr, it seems that the system age is likely even younger than the previously estimated $0.7$\,Myr  \citep{1993PASP..105..588P}. At this early stage, the primary is still a hydrogen-burning main-sequence star. 

  While the mass predicted by BONNSAI for the secondary is roughly 10\,$M_\odot$ higher than derived here, it is well within our error margin. The estimated age is $1.04\substack{+0.70\\-0.86}$\,Myr. This is a difference compared to the result for the primary. However, the error margins of the two estimates do overlap and indicate a system age  somewhere between 0.2 and 0.7\,Myr. A more drastic cause for the difference might be that these two stars are not bound to each other and their proximity is just a projection effect. Following \cite{2015A&A...579A..99B} and \citet{2017MNRAS.464.3561M}, this explanation can be ruled out. It is also possible that these stars did not evolved coevally and formed a binary system as a result of dynamical interaction.  This scenario would also be in line with orbital properties, such as the high orbital eccentricity \citep[cf.][]{2017MNRAS.464.3561M}.

  The tracks calculated by \citet{2012A&A...537A.146E} account for higher initial masses than those published by \citet{2011A&A...530A.115B}, while the Ekstr\"om grid is coarser in initial mass.  Given the young age of the two components, it is impossible to tell whether the tracks with rotation fit better than those without. However, judging by the relatively low $v_\text{rot} \sin i$ values derived in this study, the evolution of the stars is probably described more accurately by the tracks without rotation. On the other hand, we found CNO abundances in the atmospheres that indicate rotationally induced mixing. From the qualitative comparison in Fig.\,\ref{fig:hrd}, we estimate that the initial mass of the primary lies somewhere between 85$\,M_\odot$ and 120$\,M_\odot$, likely around $110\,M_\odot$. 
 
\subsection*{Is HD\,93129A a triple system?}\label{sec:triple}

  The recent study by \citet[][ MA17 hereafter]{2017MNRAS.464.3561M}  made use of sophisticated sub-resolution reduction techniques to extract separated spectra for Aa and Ab from optical HST data. MA17 discussed the hypothesis that HD\,93129A may in fact be a triple system, where the star treated here as primary (Aa) is itself a binary (Aa1 + Aa2). While Aa1 and Ab are suggested to be both of spectral type O2\,I, Aa2 probably is of slightly later subtype. MA17 presented three arguments for their three-component hypothesis, which we briefly repeat and address in the following in the light of our study.

  \begin{enumerate}
    \item The disentangled spectrum for the primary component shows line features that are inconsistent with a single star spectrum  (MA17).

      While the spectra extracted by MA17 using sophisticated sub-resolution techniques are suggestive of three components, we need to consider them with caution. It is our impression that the disentangled spectra for the Aa and Ab component presented by MA17 show artifacts that might indicate the limitations of sub-resolution extraction techniques. For example, the H$\alpha$ line extracted by MA17 for Aa shows a round emission profile, while that of Ab displays a peculiar round absorption profile of a similar width superimposed to a narrow emission (see figure\,2 in MA17). This narrow emission would imply much lower terminal velocities than derived here from the UV lines, and also lower than expected for O-type stars. These profiles potentially indicate cross-contamination of both sources. 
      
    \item Based on the orbital solution, a mass ratio of 0.5 between Ab and Aa follows. This is inconsistent with the spectral types (both are O2\,I) derived from the separated spectra. The additional mass is therefore attributed to the hypothetical third component Aa2 (MA17).

      In our analysis, the spectral types are O2\,If* and O3\,III(f) for Aa and Ab, respectively. The mass ratio of 0.5 is consistent with our spectroscopic and evolutionary mass determinations. 

    \item The spectral lines in the integrated spectrum show short-term velocity variations that are inconsistent with the long-term variations expected from the Aa Ab orbit (MA17).

      If a radial velocity curve for the Aa1 and Aa2 orbit can be extracted from these variations, this would be a convincing argument for the triple system. No such orbital solution has been  established as yet, however. As a caveat, it may well be that these radial velocity shifts are manifestations of stochastic wind instabilities, as is often observed in massive stars.
  \end{enumerate}
  Another problem of the three-component hypothesis concerns the brightness of the components. MA17 suggested that Aa1 and Ab have the same spectral type. Therefore, their brightness should be roughly identical. Given that all components add together to an absolute visual brightness of $M_\text{V} = -6.5\,$mag and that the difference between Aa and Ab was measured to $\Delta m_\text{V} = 0.9\,$mag (see Sect.\,\ref{sec:analysis}), it follows that Aa1 and Ab would each contribute $-5.2\,$mag, while Aa2 contributes another $-5.5\,$mag. In other words, the hypothetical Aa2 star should be the brightest component in the system. This seems unlikely, however, since the absorption features in the composed Aa spectrum (see figure 2 in MA17) are only weak. Additionally, the fact that the absorption troughs of the UV P-Cygni resonance profiles are saturated implies that all components of the system would have strong winds with similar terminal velocities. Based on our two component analysis, we find no obvious need to invoke a third compoment to reproduce the observations. 

  Based on radio emission measurements, \cite{2015A&A...579A..99B} estimated a ratio of $\approx 2$ between the mass-loss rates of Aa and Ab. If a similar spectral type were assumed for Aa1 and Ab, this would imply similar mass-loss rates for all three components. The radio map by \cite{2015A&A...579A..99B} clearly shows the wind-wind collision region between Aa and Ab, but no emission that could be attributed to interactions between Aa1 and Aa2. \cite{2011MNRAS.415.3354C} attributed the observed X-ray emission to intrinsic wind shocks without invoking a contribution from colliding winds.

  Given the issues with the spectral disentangled presented by MA17 as discussed above, the total spectrum of HD\,93129A might still be analyzed by adding the synthetic spectra from three components. However, since the two-component fit presented here reproduces the data at a satisfactory level, the remaining residuals do not provide enough information for an analysis with three components. If future observations were to provide radial velocity measurements that prove the binary nature of Aa, however, disentangling the component spectra might become feasible.

\section{Summary}

  HD\,93129A is the earliest spectral type binary star known in the Milky Way and served as the prototype for the spectral class O2\,If*. However, this spectral class was assigned before it was revealed that this object is a multiple system. A spectral disentanglement is required to place constraints on the spectral type and determine the stellar content of the binary. Because the orbital period is very long, radial velocity variations cannot help to  disentangle the component spectra.

  We performed the first spectral analysis of HD\,93129A that accounts for its binarity. Synthetic spectra were calculated using the Potsdam Wolf-Rayet model atmosphere code for the primary and secondary, and their combination was then compared to newly obtained high-resolution spectra from the HST in the UV and the VLT in the optical. Furthermore, all previously obtained UV data for this source were proven to be contaminated by neighboring sources. 

  We achieved a consistent spectral fit from the FUV to NIR and were able to determine the stellar parameters for both components. The synthetic spectra for the individual components allowed us to spectroscopically classify the primary as O2~If*/WN5 and the secondary as O3.5~III. 

  We find a good agreement between stellar evolution models and our derived parameters. According to these model, the system age is estimated to be $\approx 0.5\,$Myr. This system is found to be less extreme than previously assumed. Nevertheless, it provides an exceptionally strong feedback in the form of ionizing photons and mechanical energy. 

  Finally, we discussed the study by \citet{2017MNRAS.464.3561M}, who argued that the system may be a triple. These results clearly warrant further investigations that might unequivocally establish the presence or absence of additional components in the system. 

\begin{acknowledgements} 
  We thank the anonymous referee for helpful comments and suggestions.
    \\
  Based on observations made with NASA/ESA Hubble Space Telescope, obtained from the Mikulsky Archive at Space Telescope Science Institute, operated by the Association of Universities for Research in Astronomy, Inc., under NASA contract NAS 5-26555. Support for ASTRAL is provided by grants HST-GO-12278.01-A and HST-GO-13346.01-A from STScI. (PI: Ayres)
    \\
  The project has made use of public databases hosted by SIMBAD, maintained by CDS, Strasbourg, France. 
    \\
  Based on observations made with ESO telescopes at the La Silla Paranal Observatory under programme ID 095.D-0234(A). 
    \\
  This publication makes use of data products from the Two Micron All Sky Survey, which is a joint project of the University of Massachusetts and the Infrared Processing and Analysis Center/California Institute of Technology, funded by the National Aeronautics and Space Administration and the National Science Foundation.
    \\
  This research has made use of the NASA/ IPAC Infrared Science Archive, which is operated by the Jet Propulsion Laboratory, California Institute of Technology, under contract with the National Aeronautics and Space Administration.
    \\
  A.A.C.S. is supported by the Deutsche Forschungsgemeinschaft (DFG) under grant HA 1455/26 and would like to thank
STFC for funding under grant number ST/R000565/1. T.S. acknowledges support from the German “Verbundforschung” (DLR) grant 50 OR 1612. L.M.O. acknowledges support by the DLR grant 50 OR 1508.
    \\
  Some of the data presented in this paper were obtained from the Mikulski Archive for Space Telescopes (MAST). STScI is operated by the Association of Universities for Research in Astronomy, Inc., under NASA contract NAS5-26555. Support for MAST for non-HST data is provided by the NASA Office of Space Science via grant NNX09AF08G and by other grants and contracts.
    \\
  Based on observations made with the NASA/ESA Hubble Space Telescope, and obtained from the Hubble Legacy Archive, which is a collaboration between the Space Telescope Science Institute (STScI/NASA), the European Space Agency (ST-ECF/ESAC/ESA) and the Canadian Astronomy Data Centre (CADC/NRC/CSA).
    \\
  This work has made use of data from the European Space Agency (ESA) mission {\it Gaia} (\url{https://www.cosmos.esa.int/gaia}), processed by the {\it Gaia} Data Processing and Analysis Consortium (DPAC,\url{https://www.cosmos.esa.int/web/gaia/dpac/consortium}). Funding for the DPAC has been provided by national institutions, in particular the institutions participating in the {\it Gaia} Multilateral Agreement.
\end{acknowledgements}

  \bibliographystyle{aa}
  \bibliography{lit.bib}

\end{document}